\documentclass[a4paper,12pt,twoside]{article}
\usepackage{psfig}
\usepackage{epsfig}
\usepackage{amssymb,wasysym,bbm,feynmf,rotating,subfigure,here,citesort}
\usepackage{graphicx}
\newcommand{\alphaS}{\alpha_s}
\newcommand{\alphaEM}{\alpha}
\newcommand{\SVM}{SVM} 
\newcommand{\pert}{P}
\newcommand{\nprt}{N\!P}

\newcommand{\qbar}{{\bar{q}}}

\newcommand{\be}{\begin{equation}}
\newcommand{\ee}{\end{equation}}
\newcommand{\bea}{\begin{eqnarray}}
\newcommand{\eea}{\end{eqnarray}}
\newcommand{\benn}{\begin{displaymath}}
\newcommand{\eenn}{\end{displaymath}}
\newcommand{\beann}{\begin{eqnarray*}}
\newcommand{\eeann}{\end{eqnarray*}}
\newcommand{\inv}{\frac{1}}
\newcommand{\gtsim}{\lower-0.45ex\hbox{$>$}\kern-0.77em\lower0.55ex\hbox{$\sim$}}
\newcommand{\ltsim}{\lower-0.45ex\hbox{$<$}\kern-0.77em\lower0.55ex\hbox{$\sim$}}
\newcommand{\fm}{\mbox{fm}}
\newcommand{\mb}{\mbox{mb}}

\newcommand{\GeV}{\mbox{GeV}}

\newcommand{\G}{{\cal G}}       
\newcommand{\Pc}{{\cal P}}      
\newcommand{\Tr}{\mbox{Tr}}    
\newcommand{\im}{\mbox{Im}}    
\newcommand{\soft}{\mbox{\scriptsize soft}}
\newcommand{\hard}{\mbox{\scriptsize hard}}
\newcommand{\pt}{\mbox{\scriptsize pt}}
\newcommand{\befig}{\begin{figure}}
\newcommand{\efig}{\end{figure}}
\newcommand{\betab}{\begin{table}}
\newcommand{\etab}{\end{table}}
\begin{document}
%
%
%
\pagestyle{empty}
%
%
\title{ \vspace*{-1cm} {\normalsize\rightline{HD-THEP-02-21}}
  {\normalsize\rightline{hep-ph/0207287}}  \vspace*{0.cm} {\Large \bf
    \boldmath Decomposition of the QCD String into Dipoles and
  Unintegrated Gluon Distributions}}
\author{}
\date{} \maketitle
 
\vspace*{-2.5cm}
 
\begin{center}
 
\renewcommand{\thefootnote}{\alph{footnote}}
 
{\large
A.~I.~Shoshi$^{1,}$\footnote{shoshi@tphys.uni-heidelberg.de},
F.~D.~Steffen$^{1,}$\footnote{Frank.D.Steffen@thphys.uni-heidelberg.de}, 
H.G.~Dosch$^{1,}$\footnote{H.G.Dosch@thphys.uni-heidelberg.de}, and
H.~J.~Pirner$^{1,2,}$\footnote{pir@tphys.uni-heidelberg.de}}
 
 
{\it $^1$Institut f\"ur Theoretische Physik, Universit\"at Heidelberg,\\
Philosophenweg 16 {\sl \&}\,19, D-69120 Heidelberg, Germany}
 
 
{\it $^2$Max-Planck-Institut f\"ur Kernphysik, Postfach 103980, \\
D-69029 Heidelberg, Germany}
 
 
\end{center}
 

\begin{abstract}
  
  We present the perturbative and non-perturbative QCD structure of
  the dipole-dipole scattering amplitude in momentum space. The
  perturbative contribution is described by two-gluon exchange and the
  non-perturbative contribution by the stochastic vacuum model which
  leads to confinement of the quark and antiquark in the dipole via a
  string of color fields. This QCD string gives important
  non-perturbative contributions to high-energy reactions. A new
  structure different from the perturbative dipole factors is found in
  the string-string scattering amplitude. The string can be
  represented as an integral over stringless dipoles with a given
  dipole number density.  This decomposition of the QCD string into
  dipoles allows us to calculate the unintegrated gluon distribution
  of hadrons and photons from the dipole-hadron and dipole-photon
  cross section via $|\vec{k}_{\!\perp}|$\,-\,factorization.

  \vspace{1.cm}
 
\noindent
{\it Keywords}:
Confining String,
Gluon Distribution,
High-Energy Scattering,
Non-Perturbative QCD,
Stochastic Vacuum Model,
Two-Gluon Exchange
 
\medskip

\noindent
{\it PACS numbers}:
11.80.Fv,      
12.38.-t,      
12.40.-y,      
13.60.-r,      

\end{abstract}

%
%
%

\pagenumbering{roman}
\pagestyle{plain}
%
%
%
\pagenumbering{arabic}
\pagestyle{plain}
%
\makeatletter
\@addtoreset{equation}{section}
\makeatother
\renewcommand{\theequation}{\thesection.\arabic{equation}}

\newpage
\section{Introduction}
\label{Sec_Introduction}

One of the challenges in quantum chromodynamics (QCD) is the
understanding of hadronic high-energy reactions. Asymptotic freedom
allows perturbative calculations of hadron-hadron interactions at
large momentum transfer. Computer simulations on Euclidean lattices
give access to non-perturbative QCD and allow to investigate static
properties of hadrons such as the hadron spectrum and confinement.
Less is known about the non-perturbative structure of
hadron-hadron scattering as non-perturbative methods working in
Minkowski space-time are required.

In this work, we investigate the perturbative and non-perturbative QCD
structure of the dipole-dipole scattering amplitude in momentum space.
We use the recently presented loop-loop correlation model
(LLCM)~\cite{Shoshi:2002in} which combines perturbative and
non-perturbative QCD contributions and allows to compute cross
sections of hadron-hadron, photon-hadron and photon-photon reactions
in a unified description. The model is based on a functional integral
approach developed for parton-parton scattering in the eikonal
approximation~\cite{Nachtmann:1991ua,Nachtmann:ed.kt} and extended to
gauge-invariant loop-loop
scattering~\cite{Kramer:1990tr,Dosch:1994ym,Dosch:RioLecture}.  The
$S$-matrix factorizes into the universal correlation of two light-like
Wegner-Wilson loops and reaction-specific wave functions. The
light-like Wegner-Wilson loops describe color-dipoles given by the
quark and antiquark in the meson or photon and in a simplified picture
by a quark and diquark in the baryon. The size and orientation of the
color-dipoles in the hadrons and photons are determined by appropriate
light-cone wave functions. The perturbative correlation is described
by lowest order gluon exchange and the non-perturbative correlation by
the stochastic vacuum model (SVM)~\cite{Dosch:1987sk+X}. The
unitarization of both contributions obtained in the approach of Berger
and Nachtmann~\cite{Berger:1999gu} is important for ultra-high
energies and non-forward scattering~\cite{Shoshi:2002in} and is not
discussed here.

The SVM gives confinement for non-Abelian gauge theories due to
flux-tube formation of color fields between the quark and antiquark of
a
dipole~\cite{DelDebbio:1994zn,Rueter:1995cn,Euclidean_Model_Applications}.
This flux-tube or {\em string} enters via the essential SVM assumption
that non-perturbative interactions are most adequately described in
terms of correlators of gluon field strengths instead of gluon
potentials as in perturbative QCD. Therefore, the line integrals over
the gluon potentials in the Wegner-Wilson loops of the functional
integral
approach~\cite{Nachtmann:1991ua,Nachtmann:ed.kt,Kramer:1990tr,Dosch:1994ym,Dosch:RioLecture}
are transformed into surface integrals over gluon field strengths
using the non-Abelian Stokes
theorem~\cite{Bralic:1980ra+X,Nachtmann:ed.kt}.  In line with our
recent work~\cite{Shoshi:2002in} we use {\em minimal surfaces}, planar
surfaces bounded by Wegner-Wilson loops, which are usually used to
obtain Wilson's area law in Euclidean
space-time~\cite{Dosch:1987sk+X,Euclidean_Model_Applications}. Minimal
surfaces allow us to show for the first time the non-perturbative
structure of the dipole-dipole scattering amplitude in momentum space.
The previous pyramid mantle choice for the
surfaces~\cite{Dosch:1994ym,Berger:1999gu,Rueter:1996yb+X,Dosch:1997ss,Dosch:1998nw,Rueter:1998up,D'Alesio:1999sf,Dosch:2001jg}
did not allow such an analysis.

The perturbative contribution to dipole-dipole scattering in our
approach is the known two-gluon exchange~\cite{Low:1975sv+X,Gunion:iy}
between the two dipoles. The non-perturbative contribution comes out
as a sum of two parts: The first part describes the non-perturbative
interaction between the quarks and antiquarks of the two dipoles and
exhibits the same dipole factors as the perturbative contribution. The
second part represents the interaction between the strings of the two
dipoles. It shows a new structure different from the perturbative
two-gluon exchange. Nevertheless, we would like to empasize that
$|\vec{k}_{\!\perp}|$\,-\,factorization is valid also for the
non-perturbative contribution.

The string confining the quark-antiquark dipole of length
$|\vec{r}_{\!\mbox{\tiny\it D}}|$ has an integral representation which
sums stringless dipoles of sizes $\xi|\vec{r}_{\!\mbox{\tiny\it D}}|$
with $0 \leq \xi \leq 1$ and dipole number density $n(\xi) = 1/\xi^2$.
Consequently, the string-hadron scattering reduces to an incoherent
superposition of dipole-hadron scattering processes.  This
decomposition of the string into stringless dipoles allows us to
extract the microscopic structure of the unintegrated gluon
distribution of hadrons and photons, ${\cal
  F}_{\!h}(x,k_{\!\perp}^2)$, from our dipole-hadron and dipole-photon
cross section via $|\vec{k}_{\!\perp}|$\,-\,factorization.

The unintegrated gluon distribution of hadrons and photons ${\cal
  F}_{h}(x,k_{\!\perp}^2)$ is a basic, universal
quantity convenient for the computation of many scattering observables
at small $x$. It is crucial to describe processes in which transverse
momenta are explicitly exposed such as dijet~\cite{Nikolaev:1994cd+X}
or vector meson~\cite{Nemchik:1997xb} production at HERA. Its explicit
$|\vec{k}_{\!\perp}|$\,-\,dependence is particularly suited to study the
interplay between soft and hard physics. Moreover, the unintegrated
gluon distribution is the central object in the
BFKL~\cite{Kuraev:fs+X} and CCFM~\cite{Ciafaloni:1987ur+X} evolution
equations. Upon integration over the transverse gluon momentum
$|\vec{k}_{\!\perp}|$ it leads to the conventional gluon distribution
$xG_{h}(x,Q^2)$ used in the DGLAP evolution
equation~\cite{Gribov:ri+X}. 

The outline of the paper is as follows: In
Sec.~\ref{Sec_The_Loop_Loop_Correlation_Model}, we review the
loop-loop correlation model~\cite{Shoshi:2002in} and give the model
parameters. In Sec.~\ref{Momentum-Space Structure of Dipole-Dipole
  Scattering}, we investigate the perturbative and non-perturbative
structure of the dipole-dipole scattering amplitude in momentum space.
In
Sec.~\ref{Sec_The_Decomposition_of_the_String_into_Dipoles_and_the_Unintegrated_Gluon_Distribution},
we present the decomposition of the string into dipoles, elaborate the
total dipole-hadron cross section, and extract the unintegrated gluon
distribution. Numerical results for the unintegrated gluon
distribution of protons, pions, kaons, and photons are given in
Sec.~\ref{Sec_Numerical_Results_of_Unintegrated_Gluon_Distributions_of_Hadrons_and_Photons}
together with the integrated gluon distribution of the proton
$xG_p(x,Q^2)$. In Sec.~\ref{Sec_Comparison_with_Other_Work}, we
compare our results for the unintegrated gluon distribution of the
proton with those obtained in other approaches. We close with a
summary of our results.

\section{The Loop-Loop Correlation Model}
\label{Sec_The_Loop_Loop_Correlation_Model}

Recently, we have presented a loop-loop correlation model to compute
high-energy hadron-hadron, photon-hadron, and photon-photon
reactions~\cite{Shoshi:2002in}. Based on the functional integral
approach to high-energy scattering in the eikonal
approximation~\cite{Nachtmann:1991ua,Nachtmann:ed.kt,Kramer:1990tr,Dosch:1994ym,Dosch:RioLecture},
the $T$-matrix element for the elastic scattering of two particles $1$
and $2$ at transverse momentum transfer ${\vec q}_{\!\perp}$ ($t =
-{\vec q}_{\!\perp}^{\,\,2}$) and c.m.\ energy squared~$s$ reads
\bea
        \!\!\!\!\!\!
        T(s,t) &=&
        2is \int \!\!d^2b_{\!\perp} 
        e^{i {\vec q}_{\!\perp} {\vec b}_{\!\perp}}
        \int \!\!dz_1 d^2r_1 \!\int \!\!dz_2 d^2r_2        
        \nonumber \\ 
        & & \times\, 
        |\psi_1(z_1,{\vec r}_1)|^2\,|\psi_2(z_2,{\vec r}_2)|^2
        \left[1-S_{DD}({\vec b}_{\!\perp},z_1,{\vec r}_1,z_2,{\vec r}_2)\right]
        \ , 
\label{Eq_model_T_amplitude}
\eea
where the correlation of two light-like Wegner-Wilson loops, the {\em
  loop-loop correlation function},\footnote{More generally, $S_{DD} =
  \langle W[C_1] W[C_2] \rangle_G / (\langle W[C_1]\rangle_G \langle
  W[C_2]\rangle_G)$. For light-like Wegner-Wilson loops, however, one
  obtains $\langle W[C_{1,2}] \rangle_G = 1$ as shown explicitly
  in~\cite{Nachtmann:ed.kt}.}
\be
        S_{DD}({\vec b}_{\!\perp},z_1,{\vec r}_1,z_2,{\vec r}_2)
        = \Big\langle W[C_1] W[C_2] \Big\rangle_G \ ,
\label{Eq_loop_loop_correlation_function}
\ee
describes the elastic scattering of two color-dipoles (DD).  The
path of each color-dipole is represented by a light-like Wegner-Wilson
loop~\cite{Wilson:1974sk+X} in the fundamental representation of
$SU(3)$
\be
        W[C_{i}] = 
        \inv{3} \, \Tr \ \Pc
        \exp\!\left[-i g\!\oint_{\mbox{\scriptsize $C_{i}$}}\!\!dz^{\mu}
        \G_{\mu}(z) \right]        
        \ ,
\label{Eq_Wegner-Wilson_loop}
\ee
where $\Tr$ is the trace in color space, $g$ the strong coupling, and
$\G_{\mu}(z) = \G_{\mu}^a(z) t^a$ the gluon field with the $SU(3)$
group generators $t^a$ that demand the path ordering indicated by
$\Pc$.  Physically, the Wegner-Wilson loops
(\ref{Eq_Wegner-Wilson_loop}) represent the phase that quarks and
antiquarks acquire along the light-like trajectories $C_{i}$ in the
gluon background field. The color-dipoles have transverse size and
orientation ${\vec r}_i$. The longitudinal momentum fraction of dipole
$i$ carried by the quark is $z_i$. The impact parameter between the
dipoles is~\cite{Dosch:1997ss}
\be
        {\vec b}_{\!\perp} 
        \,=\, {\vec r}_{1q} + (1-z_1) {\vec r}_{1} 
            - {\vec r}_{2q} - (1-z_2) {\vec r}_{2} 
        \,=\, {\vec r}_{1\,cm} - {\vec r}_{2\,cm} 
        \ ,
\label{Eq_impact_vector}
\ee
where ${\vec r}_{iq}$ (${\vec r}_{i\qbar}$) is the transverse position
of the quark (antiquark), ${\vec r}_{i} = {\vec r}_{i\qbar} - {\vec
  r}_{iq}$, and ${\vec r}_{i\,cm} = z_i {\vec r}_{iq} + (1-z_i){\vec
  r}_{i\qbar}$ is the center of light-cone momenta.
\befig[p!]
  \begin{center}
        \epsfig{file=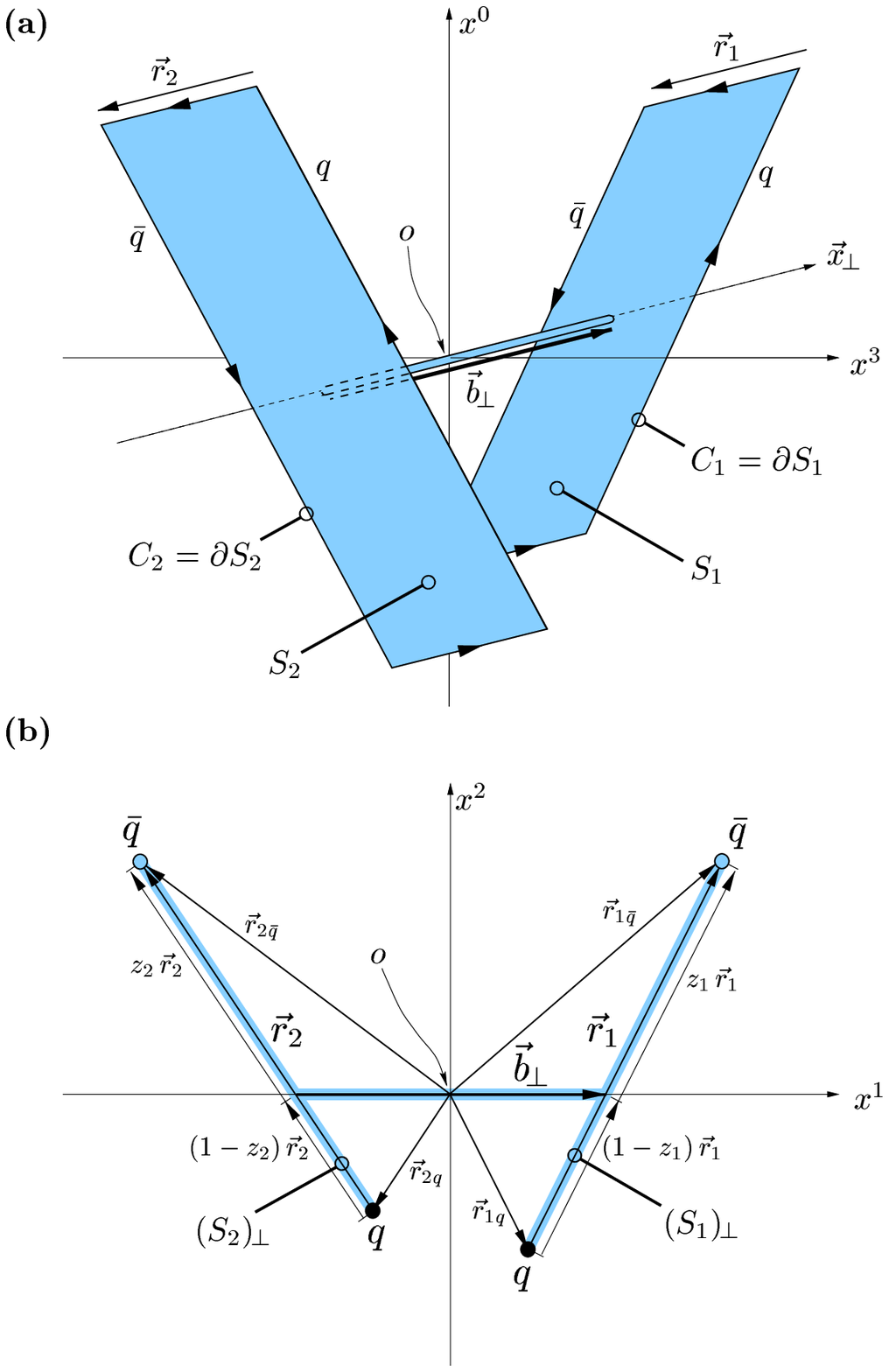,width=10.cm}
  \end{center}
\caption{\small High-energy dipole-dipole scattering in the eikonal
  approximation represented by Wegner-Wilson loops: (a) space-time and
  (b) transverse arrangement of the Wegner-Wilson loops. The shaded
  areas represent the strings extending from the quark to the
  antiquark path in each color dipole.  The thin tube allows to
  compare the field strengths in surface $S_1$ with the field
  strengths in surface $S_2$. The impact parameter $\vec{b}_{\perp}$
  connects the centers of light-cone momenta of the dipoles.}
\label{Fig_loop_loop_scattering_surfaces}
\efig
Figure~\ref{Fig_loop_loop_scattering_surfaces} illustrates the (a)
space-time and (b) transverse arrangement of the loops. The QCD vacuum
expectation value $\langle \ldots \rangle_G$ in the loop-loop
correlation function (\ref{Eq_loop_loop_correlation_function})
represents functional integrals~\cite{Nachtmann:ed.kt} in which the
functional integration over the fermion fields has already been
carried out as indicated by the subscript $G$.  The model we use for
the QCD vacuum describes only gluon dynamics and, thus, implies the
quenched approximation that does not allow string breaking through
dynamical quark-antiquark production.

\medskip

The ${\vec r}_i$ and $z_i$ distribution of the color-dipoles is given
by the {\em wave functions} $\psi_{i}$ that characterize the
interacting particles.  In this framework, the color-dipoles are given
by the quark and antiquark in the meson or photon and in a simplified
picture by a quark and diquark in the baryon. We use for hadrons the
phenomenological Gaussian wave
function~\cite{Dosch:2001jg,Wirbel:1985ji} and for photons the
perturbatively derived wave functions with running quark masses
$m_f(Q^2)$ to account for the non-perturbative region of low photon
virtuality $Q^2$~\cite{Dosch:1998nw} as discussed explicitly in
Appendix~\ref{Sec_Wave_Functions}.

\medskip

The computation of the loop-loop correlation
function~(\ref{Eq_loop_loop_correlation_function}) in the matrix
cumulant approach of Berger and Nachtmann~\cite{Berger:1999gu} has
been reviewed in detail in~\cite{Shoshi:2002in}. The main steeps of
this computation are the transformation of the line integrals into
surface integrals with the non-Abelian Stokes theorem, a matrix
cumulant expansion, and the Gaussian approximation of the functional
integrals.  In line with other two-component (soft $+$ hard)
models~\cite{Donnachie:1998gm+X,D'Alesio:1999sf,Forshaw:1999uf,Rueter:1998up,Donnachie:2001wt}
and the different hadronization mechanisms in soft and hard
collisions, we have further demanded that the perturbative and
non-perturbative contributions do not mix on the amplitude
level~\cite{Shoshi:2002in}. The resulting $T$-matrix element at the
reference c.m.\ energy $\sqrt{s_0}$ reads
\bea
        && \!\!\!\!\!\!\!\!\!\!
        T(s_0,t) 
        = 2is_0 \int \!\!d^2b_{\!\perp} 
        e^{i {\vec q}_{\!\perp} {\vec b}_{\!\perp}}
        \int \!\!dz_1 d^2r_1 \!\int \!\!dz_2 d^2r_2\,\,
        |\psi_1(z_1,\vec{r}_1)|^2 \,\, 
        |\psi_2(z_2,\vec{r}_2)|^2       
        \nonumber \\    
        &&\!\!\!\!\!\!\!\!\!\!\!\! 
        \times \left[ 1 - \frac{2}{3} 
        \cos\left(\frac{1}{3}\chi^{\pert}\right)
        \cos\left(\frac{1}{3}\chi^{\nprt}\right)         
        - \frac{1}{3}
        \cos\left(\frac{2}{3}\chi^{\pert}\right)
        \cos\left(\frac{2}{3}\chi^{\nprt}\right)
        \right] \ ,
\label{Eq_model_purely_imaginary_T_amplitude_almost_final_result}
\eea
where the arguments of the functions $\chi^{\pert,\,\nprt}({\vec
  b}_{\!\perp},z_1,\vec{r}_1,z_2,\vec{r}_2)$ are suppressed to lighten
notation. The ${\chi}$\,-\,functions defined as
\be    
        \chi^{\pert,\,\nprt} 
        :=- \,i\,\frac{\pi^2}{4}\!
        \int_{S_1} \! d\sigma^{\mu\nu}(x_1) 
        \int_{S_2} \! d\sigma^{\rho\sigma}(x_2)\,
          F_{\mu\nu\rho\sigma}^{\pert,\,\nprt}(x_1,x_2,o;C_{x_1 o},C_{x_2 o})
        \ 
\label{chi_amplitude}
\ee
represent integrals over the {\em minimal surfaces} $S_{1,2}$ that are
built from the areas spanned by the quark and antiquark paths
$C_{1,2}$ of the interacting dipoles and the infinitesimally thin tube
which connects the surfaces $S_1$ and $S_2$ as shown in
Fig.~\ref{Fig_loop_loop_scattering_surfaces}.
The integrand $F_{\mu\nu\rho\sigma}^{\pert,\,\nprt}$ 
is the gauge-invariant bilocal gluon field strength correlator 
\be
\Big\langle \frac{g^2}{4\pi^2}\, \G^a_{\mu\nu}(o,x_1;C_{x_1 o})
\G^b_{\rho\sigma}(o,x_2;C_{x_2 o})
\Big\rangle_G^{\pert,\,\nprt}=:\inv{4}\delta^{ab}
F_{\mu\nu\rho\sigma}^{\pert,\,\nprt}\!  (x_1,x_2,o;C_{x_1
  o},C_{x_2 o}) \ ,
\label{Eq_Ansatz}
\ee
where the gluon field strengths $\G_{\mu\nu}(x_i)$ are parallel
transported to the common reference point $o$ along the path
$C_{x_io}$
\be
        \G_{\mu\nu}(o,x_i;C_{x_io}) 
        = \Phi(x_i,o;C_{x_io})^{-1} \G_{\mu\nu}(x_i) \Phi(x,o;C_{x_io})
\label{Eq_gluon_field_strength_tensor}
\ee
with the QCD Schwinger string
\be
        \Phi(x_i,o;C_{x_io}) 
        = \Pc \exp 
        \left[-i g \int_{C_{x_io}} dz^{\mu}\G_{\mu}(z)\right] 
\label{Eq_parallel_transport}
\ee
to ensure gauge invariance in the model. In~(\ref{Eq_Ansatz}),
$F_{\mu\nu\rho\sigma}^{\pert}$ gives the perturbative ($\pert$)
physics (short-range correlations) described by {\em perturbative
  gluon exchange} and $F_{\mu\nu\rho\sigma}^{\nprt}$ the
non-perturbative ($\nprt$) physics (long-range correlations) modelled
by the {\em stochastic vacuum model} (SVM)~\cite{Dosch:1987sk+X}.

In leading order in the strong coupling $g$, the perturbative bilocal
gluon field strength correlator $F_{\mu\nu\rho\sigma}^{\pert}$ is
gauge-invariant already without the parallel transport to a common
reference point and depends only on the difference $z:= x_1 - x_2$,
\bea
        F_{\mu\nu\rho\sigma}^{\pert}(z)
        \!\!&=&\!\!\frac{g^2}{\pi^2}\, \inv{2}\Bigl[
                       \frac{\partial}{\partial z_\nu}
                         \left(z_\sigma g_{\mu\rho}
                         -z_\rho g_{\mu\sigma}\right)
                       +\frac{\partial}{\partial z_\mu}
                         \left(z_\rho g_{\nu\sigma}
                         -z_\sigma g_{\nu\rho}\right)\Bigr]\,
              D_{\pert}(z^2)
        \\
        \!\!&=&\!\! -\,\frac{g^2}{\pi^2}\!
                \int \!\!\frac{d^4k}{(2\pi)^4} \,e^{-ikz}\,\Bigl[
                k_\nu k_\sigma g_{\mu\rho}  - k_\nu k_\rho   g_{\mu\sigma}
              + k_\mu k_\rho  g_{\nu\sigma} - k_\mu k_\sigma g_{\nu\rho} \Bigr]\,
           \tilde{D}_{\pert}^{\prime}(k^2) 
        \nonumber \ .
\label{Eq_PGE_Ansatz_F}
\eea
Here, we have introduced an effective gluon mass $m_G = m_{\rho} = 0.77\,\GeV$ as infrared regulator in the
perturbative correlation function 
\be
        \tilde{D}_{\pert}^{\prime}(k^2) 
        := \frac{d}{dk^2} \int d^4z \,D_{\pert}(z^2)\, e^{ikz} 
         = \frac{i}{k^2 - m_G^2} \ ,
\label{Eq_massive_D_pge_prime}
\ee
and a parameter $M^2 = 1.04\,\GeV^2$ which freezes the running
coupling\footnote{Only the running coupling as a function of the
  transverse momentum is needed in the final result.} in the quenched
approximation at the value $\alpha_s(k_{\!\perp}^2=0) =
0.4$~\cite{Euclidean_Model_Applications},
\be
        \alpha_s(k_{\!\perp}^2) = \frac{g^2(k_{\!\perp}^2)}{4\pi}
        = \frac{1}
        {11 \ln\left[(k^2_{\!\perp} + M^2)/\Lambda_{QCD}^2
        \right]} \ .
\label{Eq_g2(z_perp)}
\ee

If the path connecting the points $x_1$ and $x_2$ is a straight line,
the non-perturbative correlator $F_{\mu\nu\rho\sigma}^{\nprt}$ depends
also only on the difference $z:= x_1 - x_2$. Then, the most general
form of the correlator in four-dimensional Minkowski space-time that
respects translational, Lorentz, and parity invariance
reads~\cite{Kramer:1990tr,Dosch:1994ym,Dosch:RioLecture}
\bea
        \!\!\!\!
      F_{\mu\nu\rho\sigma}^{\nprt}(z) 
        &\!\!\!:=\!\!\!& F_{\mu\nu\rho\sigma}^{\nprt_{(nc)}}(z) +  
               F_{\mu\nu\rho\sigma}^{\nprt_{(c)}}(z)
\label{Eq_MSV_Ansatz_F}\\
\vspace*{-1cm}\nonumber\\
        \!\!\!\!
        F_{\mu\nu\rho\sigma}^{\nprt_{(nc)}}(z)
        &\!\!\!=\!\!\!&  \frac{G_2\,(1\!-\!\kappa)}{6(N_c^2-1)} 
        \Bigl(\frac{\partial}{\partial z_\nu}
        \left(z_\sigma g_{\mu\rho}
          \!-\!z_\rho g_{\mu\sigma}\right)
        \!+\!\frac{\partial}{\partial z_\mu}
        \left(z_\rho g_{\nu\sigma}
          \!-\!z_\sigma g_{\nu\rho}\right)\Bigr)\,
        D_1(z^2) 
\label{Eq_MSV_Ansatz_F_nc}\\
        && \hspace{-2.cm}=-\,\frac{G_2\,(1-\kappa)}{6(N_c^2-1)} 
        \int \frac{d^4k}{(2\pi)^4} \,e^{-ikz} \Bigl(
        k_\nu k_\sigma g_{\mu\rho}   - k_\nu k_\rho   g_{\mu\sigma}
        + k_\mu k_\rho  g_{\nu\sigma} - k_\mu k_\sigma g_{\nu\rho} \Bigr)\,
               \tilde{D}_{1}^{\prime}(k^2) 
\nonumber\\
\vspace*{1cm}\nonumber\\
        \!\!\!\!
        F_{\mu\nu\rho\sigma}^{\nprt_{(c)}}(z)
        &\!\!\!=\!\!\!& \frac{G_2\,\kappa}{3(N_c^2-1)}\, 
        \left(g_{\mu\rho}g_{\nu\sigma}
          -g_{\mu\sigma}g_{\nu\rho}\right) \,
        D(z^2)                                
        \nonumber\\
        &\!\!\!=\!\!\!& \frac{G_2\,\kappa}{3(N_c^2-1)}\,
        \left(g_{\mu\rho}g_{\nu\sigma}
          -g_{\mu\sigma}g_{\nu\rho}\right)\, 
        \int \frac{d^4k}{(2\pi)^4} \,e^{-ikz}\,
        \tilde{D}(k^2) 
\label{Eq_MSV_Ansatz_F_c}
\eea
and was originally constructed in Euclidean
space-time~\cite{Dosch:1987sk+X}.  In all previous applications of the
SVM, this form depending only on $x_1$ and $x_2$ has been used. New
lattice results on the path dependence of the
correlator~\cite{DiGiacomo:2002mq} show a dominance of the shortest
path. This result is effectively incorporated in the model since the
straight paths dominate in the average over all paths.

In~(\ref{Eq_MSV_Ansatz_F}), $D_1$ and $D$ are the non-perturbative
correlation functions in four-dimensional Minkowski space-time and
\be
        \tilde{D}_{1}^{\prime}(k^2) 
        := \frac{d}{dk^2} \int d^4z D_{1}(z^2) \ e^{ikz} 
        \ .
\label{Eq_D1_prime}
\ee
The Minkowskian correlation functions $D_1$ and $D$ are
derived~\cite{Shoshi:2002in} from the simple {\em exponential
  correlation functions} in four-dimensional Euclidean space-time
\be
        D^{E}_1(Z^2) = D^{E}(Z^2) = \exp(-|Z|/a)
        \ ,
\label{Eq_MSV_correlation_functions}
\ee
that are motivated by lattice QCD measurements of the gluon field
strength correlator
$F_{\mu\nu\rho\sigma}^{\nprt}(Z)$~\cite{DiGiacomo:1992df+X,Meggiolaro:1999yn}.
The parameters are the correlation length $a= 0.302\,\fm$, the gluon
condensate $G_2 := \langle \frac{g^2}{4\pi^2} \G^a_{\mu\nu}(0)
\G^a_{\mu\nu}(0) \rangle= 0.074\,\GeV^4$, and the relative weight of
the two different tensor structures $\kappa= 0.74$.

The correlator~(\ref{Eq_MSV_Ansatz_F}) is a sum of the two different
tensor structures $F_{\mu\nu\rho\sigma}^{\nprt_{(nc)}}(z)$ and
$F_{\mu\nu\rho\sigma}^{\nprt_{(c)}}(z)$. The
first~(\ref{Eq_MSV_Ansatz_F_nc}) is characteristic for Abelian gauge
theories and does not lead to confinement, while the
second~(\ref{Eq_MSV_Ansatz_F_c}) can only occur in non-Abelian gauge
theories and Abelian gauge-theories with monopoles and leads to
confinement when considered in Euclidean
space-time~\cite{Dosch:1987sk+X}. Therefore, we call the tensor
structure multiplied by $(1-\kappa)$ non-confining ($nc$) and the one
multiplied by $\kappa$ confining ($c$).

With the correlators~(\ref{Eq_PGE_Ansatz_F})
and~(\ref{Eq_MSV_Ansatz_F}) and the minimal surfaces shown in
Fig.~\ref{Fig_loop_loop_scattering_surfaces}, the functions
${\chi}^{\pert}$ and ${\chi}^{\nprt} = {\chi}^{\nprt}_{nc} +
{\chi}^{\nprt}_{c}$ have the following form~\cite{Shoshi:2002in}
\bea
\!\!\!\!\!\!
\chi^{\pert}&=&
         4\pi \int \! \frac{d^2k_{\!\perp}}{(2\pi)^2} \ 
         \alpha_s(k_{\!\perp}^2)\,i\tilde{D}_{\pert}^{\prime \,(2)}(k_{\!\perp}^2)\,
          \left [ e^{i\vec{k}_{\!\perp}( 
          \vec{r}_{1q}-\vec{r}_{2q})} + e^{i\vec{k}_{\!\perp}( 
          \vec{r}_{1\bar{q}}-\vec{r}_{2\bar{q}})} \right . 
          \nonumber \\
          &&
          \hphantom{\chi^{\pert} = 4\pi \int \! \frac{d^2k_{\!\perp}}{(2\pi)^2} \ 
          \alpha_s(k_{\!\perp}^2)\,iD} 
          \left .
          - e^{i\vec{k}_{\!\perp}( 
          \vec{r}_{1q}-\vec{r}_{2\bar{q}})} - e^{i\vec{k}_{\!\perp}( 
          \vec{r}_{1\bar{q}}-\vec{r}_{2q})} \right ]  
\label{chi_p}  \\ \vspace{2cm} \nonumber \\
\!\!\!\!\!\!
\chi_{nc}^{\nprt}&=&
         \frac{\pi^2\,G_2\,(1-\kappa)}{24}  
         \int \! \frac{d^2k_{\!\perp}}{(2\pi)^2} \ 
         i\tilde{D}^{\prime \,(2)}_{1}(k_{\!\perp}^2)\,
         \left [ e^{i\vec{k}_{\!\perp} (\vec{r}_{1q}-\vec{r}_{2q})} + 
         e^{i\vec{k}_{\!\perp} (\vec{r}_{1\bar{q}}-
         \vec{r}_{2\bar{q}})} \right . 
         \nonumber \\
         &&\!\!\!\!\!\!\!\!\!
         \hphantom{\chi_{nc}^{\nprt} = \frac{\pi^2\,G_2\,(1-\kappa)}{24} 
          \int \! \frac{d^2k_{\!\perp}}{(2\pi)^2} 
          i\tilde{D}D}
         \left .
         - e^{i\vec{k}_{\!\perp} ( 
         \vec{r}_{1q}-\vec{r}_{2\bar{q}})} - e^{i \vec{k}_{\!\perp} ( 
         \vec{r}_{1\bar{q}}-\vec{r}_{2q})} \right ]          
\label{chi_nc} \\ \vspace{2cm} \nonumber \\
\!\!\!\!\!\!
\chi_{c}^{\nprt}&=& 
       \frac{\pi^2\,G_2\,\kappa}{24}
       \left(\vec{r}_1\cdot\vec{r}_2\right) \, 
       \int_0^1 \! dv_1 \int_0^1 \! dv_2 \, \int \!
       \frac{d^2k_{\!\perp}}
       {(2\pi)^2} \ 
       i\tilde{D}^{(2)}(k_{\!\perp}^2)\, e^{i\vec{k}_{\!\perp} \left
       (\vec{r}_{1q}+ v_1\vec{r}_1-\vec{r}_{2q}-v_2\vec{r}_2\right ) }
       \ , \nonumber \\
\label{chi_c}
\eea
where $\tilde{D}_{\pert}^{\prime \,(2)}(k^2_{\!\perp})$,
$\tilde{D}^{\prime \,(2)}_1(k^2_{\!\perp})$ and
$\tilde{D}^{(2)}(k^2_{\!\perp})$ are the Minkowskian correlation functions in
transverse space obtained
from~(\ref{Eq_massive_D_pge_prime})
and~(\ref{Eq_MSV_correlation_functions}),
\bea
        i\tilde{D}_{\pert}^{\prime \,(2)}(k_{\!\perp}^2) 
         &=& \frac{1}{k^2_{\!\perp} + m_G^2} \ , 
\label{Eq_massive_D_pge_prime_trans} \\ \vspace{1cm} \nonumber \\
        i\tilde{D}^{\prime \,(2)}_1(k^2_{\!\perp}) 
        &=&  \frac{30\,\pi^2}{a(k^2_{\!\perp} +
        a^{-2})^\frac{7}{2}} \ , 
\label{Eq_D_(prime)(k^2)_for_exp_correlation}\\
        i\tilde{D}^{(2)}(k^2_{\!\perp}) 
        &=&  \frac{12\,\pi^2}{a\,(k^2_{\!\perp} +
        a^{-2})^\frac{5}{2}} \ .
\label{Eq_D(k^2)_for_exp_correlation} 
\eea
The component $\chi^{\pert}$ describes the perturbative interaction of
the quark and antiquark of one dipole with the quark and antiquark of
the other dipole as evident from the $\vec{r}_{iq}$ and
$\vec{r}_{i{\bar q}}$ dependence of~(\ref{chi_p}) and
Fig.~\ref{Fig_loop_loop_scattering_surfaces}b. The component
$\chi_{nc}^{\nprt}$ has the same structure as $\chi^{\pert}$ and gives
the non-perturbative interaction between the quarks and antiquarks of
the dipoles. With the term quark used genuinely for quarks and
antiquarks in the following, we refer to $\chi^{\pert}$ and
$\chi_{nc}^{\nprt}$ as {\em quark-quark interactions}.

The component $\chi_{c}^{\nprt}$ given in~(\ref{chi_c}) shows a
different structure. Here, the integrations over $v_1$ and $v_2$ sum
non-perturbative interactions between the gluon field strengths
connecting the quark and antiquark in each of the two dipoles. These
connections are manifestations of the strings that confine the
corresponding quark and antiquark in the dipole and are visualized in
Fig.~\ref{Fig_loop_loop_scattering_surfaces}b.  Indeed, the \SVM shows
a flux tube between a static quark-antiquark pair in Euclidean
space-time~\cite{DelDebbio:1994zn,Rueter:1995cn,Euclidean_Model_Applications}.
Therefore, we understand the confining component $\chi_{c}^{\nprt}$ as
a {\em string-string interaction}.

The mixed contribution $\chi_{nc}^{\nprt}\,\chi_{c}^{\nprt}$ that
occurs in cross sections discussed in the next sections gives the non-perturbative
interaction of the quark and antiquark of one dipole with the string
of the other dipole, i.e., it represents the {\em quark-string
  interaction}.

In our previous paper~\cite{Shoshi:2002in}, we studied extensively the
phenomenological performance of the model. With a strong and weak
powerlike energy dependence ascribed respectively to $\chi^{\pert}$
and $\chi^{\nprt}$, we have achieved a successful description of many
different observables for hadron-hadron, proton-photon and
photon-photon reactions over a wide energy range. Therefore, we adopt
from~\cite{Shoshi:2002in} all given parameter values. The
phenomenological energy dependence, however, is not essential for the
structural aspects discussed in this work. Observables obtained with the $T$-matrix
element~(\ref{Eq_model_purely_imaginary_T_amplitude_almost_final_result})
agree with experimental data at a fixed c.m.\ energy of $\sqrt{s_0}
\approx 20\,\GeV$.

In this paper, we focus on the analytic structure of the model
and concentrate on the limit of small $\chi$-functions,
$\chi^{\pert} < 1$ and $\chi^{\nprt} < 1$, in which the $T$-matrix
element~(\ref{Eq_model_purely_imaginary_T_amplitude_almost_final_result})
simplifies to the sum of a perturbative ($\pert$) and non-perturbative
($\nprt$) component
\bea
        T(s_0,t) 
        & = & 2is_0 \!\int \!\!d^2b_{\!\perp} 
        e^{i {\vec q}_{\!\perp} {\vec b}_{\!\perp}}
        \!\int \!\!dz_1 d^2r_1 \!\int \!\!dz_2 d^2r_2\,\,
        |\psi_1(z_1,\vec{r}_1)|^2 \,\,
        |\psi_2(z_2,\vec{r}_2)|^2       
        \nonumber \\  
        && \times \,\frac{1}{9}\left[
        \left(\chi^{\pert}\right)^2
        +\left(\chi^{\nprt}\right)^2
        \right].
\label{Eq_model_purely_imaginary_T_amplitude_small_chi_limit}
\eea
We show in the next section that the perturbative component,
$(\chi^{\pert})^2$, describes {\em two-gluon
  exchange}~\cite{Low:1975sv+X,Gunion:iy} and that the non-perturbative
component, $(\chi^{\nprt})^2$, represents the corresponding
non-perturbative two-point interaction. The main motivation of
this paper is to calculate the analytic structure of this
non-perturbative interaction in momentum space.
  
\section{The Structure of Dipole-Dipole Scattering in Momentum Space}
\label{Momentum-Space Structure of Dipole-Dipole Scattering}

In this section, we calculate the perturbative and non-perturbative
QCD contribution to the dipole-dipole ($DD$) scattering amplitude in
momentum space. Beyond the known perturbative two-gluon-exchange
interaction between the dipoles~\cite{Low:1975sv+X,Gunion:iy} new
insights into the non-perturbative structure of the scattering process
are obtained within the \SVM. The non-perturbative string-string
interactions show a new structure different from the dipole factors of
the perturbative contribution.

The total dipole-dipole cross section in the small-${\chi}$ limit is obtained
from~(\ref{Eq_model_purely_imaginary_T_amplitude_small_chi_limit}) via
the optical theorem
\bea
\sigma^{tot}_{\!\mbox{\tiny\it D}\mbox{\tiny\it D}}(s_0) 
            &=& \inv{s_0}\,\im\,T(s_0, t=0)
            \nonumber \\
            &=&2 \int \!\!d^2b_{\!\perp} \int \!\!dz_1d^2r_1 
            \int \!\!dz_2 d^2r_2 
            |\psi_{\!\mbox{\tiny\it D}_1}(z_1,\vec{r}_1)|^2 
            |\psi_{\!\mbox{\tiny\it D}_2}(z_2,\vec{r}_2)|^2 \nonumber \\
            && \times \frac{1}{9} 
            \left[\left(\chi^{\pert}\right)^2 +
            \left(\chi^{\nprt}_{nc} + \chi^{\nprt}_{c} \right)^2  
            \right ] \ , 
\label{sigtot_DD}
\eea  
where the dipoles have fixed $z_i$ and $|\vec{r}_i|$ values but are
averaged over all orientations 
\be
|\psi_{\!\mbox{\tiny\it D}_i}(z_i,\vec{r}_i)|^2 =
     \inv{2\pi|\vec{r}_{\!\mbox{\tiny\it D}_i}|}\,\delta
     (|\vec{r}_i|-|\vec{r}_{\!\mbox{\tiny\it D}_i}|)\,\delta
     (z_i-z_{\!\mbox{\tiny\it D}_i}) \ .
\label{dip_wf}
\ee
All integrations in~(\ref{sigtot_DD}) can be carried out analytically
except the one over transverse momentum $|\vec{k}_{\!\perp}|$ that
enters through the ${\chi}$-functions~(\ref{chi_p})--(\ref{chi_c}). The
perturbative ($\pert$) and non-perturbative ($\nprt$) integrand of the
resulting total dipole-dipole cross section
\bea
\sigma^{tot}_{\!\mbox{\tiny\it D}\mbox{\tiny\it D}}(s_0) 
     &=& 
     \int d|\vec{k}_{\!\perp}|\, \left[ 
     I^{\pert}(s_0, |\vec{k}_{\!\perp}|) +
     I^{\nprt}(s_0, |\vec{k}_{\!\perp}|)\right]
   \\
     &=& 
     \int d|\vec{k}_{\!\perp}|\, \left[ 
     I^{\pert}(s_0, |\vec{k}_{\!\perp}|) +
     I^{\nprt}_{qq}(s_0, |\vec{k}_{\!\perp}|) +
     I^{\nprt}_{ss}(s_0, |\vec{k}_{\!\perp}|)
    \right] \ ,
\label{sigt}
\eea
show the following momentum-space structure
\bea
I^{\pert}(s_0, |\vec{k}_{\!\perp}|)&=& 
      \frac{2}{9}\frac{1}{2\pi} |\vec{k}_{\!\perp}| 
      \left (4\pi \alphaS(k^2_{\!\perp})\right)^2 
      \left [i\tilde{D}_{\pert}^{\prime \,(2)}(k^2_{\!\perp})\right ]^2
\label{Ip} \\
      &&\times
      \left [2\ \langle \psi_{\!\mbox{\tiny\it D}_1}|1-e^{i\vec{k}_{\!\perp}\vec{r}_1}
      |\psi_{\!\mbox{\tiny\it D}_1} \rangle \ 
      2\ \langle \psi_{\!\mbox{\tiny\it D}_2}|1-e^{i\vec{k}_{\!\perp}\vec{r}_2}
      |\psi_{\!\mbox{\tiny\it D}_2} \rangle \right ] 
\nonumber \\
I^{\nprt}_{qq}(s_0, |\vec{k}_{\!\perp}|)
      &=& 
      \frac{2}{9}\frac{1}{2\pi} |\vec{k}_{\!\perp}| 
      \left(\frac{\pi^2 G_2}{24}\right)^2
      \left [ 
        (1-\kappa)\,i \tilde{D}^{\prime \,(2)}_1(k^2_{\!\perp})\!
        + \frac{\kappa}{k_{\!\perp}^2} \,i\tilde{D}^{(2)}(k^2_{\!\perp})
      \right]^2
\label{Eq_I_NP_qq}
      \\
      &&\times
      \left [2\ \langle \psi_{\!\mbox{\tiny\it D}_1}
      |1-e^{i\vec{k}_{\!\perp}\vec{r}_1} 
      |\psi_{\!\mbox{\tiny\it D}_1}\rangle \  
      2\ \langle \psi_{\!\mbox{\tiny\it D}_2}
      |1-e^{i\vec{k}_{\!\perp}\vec{r}_2} 
      |\psi_{\!\mbox{\tiny\it D}_2}\rangle\right]
      \nonumber\\
I^{\nprt}_{ss}(s_0, |\vec{k}_{\!\perp}|)
      &=& 
      \frac{2}{9}\frac{1}{2\pi} |\vec{k}_{\!\perp}| 
      \left(\frac{\pi^2 G_2}{24}\right)^2
      \left[\frac{\kappa}{k_{\!\perp}^2} \,i\tilde{D}^{(2)}(k^2_{\!\perp})\right]^2 
\label{Eq_I_NP_ss}
      \\
      &&\times
      \left [2\ \langle \psi_{\!\mbox{\tiny\it D}_1}|\tan^2\!\!\phi_1 
      (1-e^{i\vec{k}_{\!\perp}\vec{r}_1})
      |\psi_{\!\mbox{\tiny\it D}_1}\rangle \  
      2\ \langle \psi_{\!\mbox{\tiny\it D}_2}|\tan^2\!\!\phi_2 
      (1-e^{i\vec{k}_{\!\perp}\vec{r}_2}) 
      |\psi_{\!\mbox{\tiny\it D}_2}\rangle \right]
      \nonumber
\eea
where the brackets $\langle \psi_{\!\mbox{\tiny\it D}_i}| ...
|\psi_{\!\mbox{\tiny\it D}_i}\rangle$ denote the averages
\be
\langle \psi_{\!\mbox{\tiny\it D}_i}|A_i|\psi_{\!\mbox{\tiny\it D}_i}\rangle = 
      \int \!dz_id^2r_i |\psi_{\!\mbox{\tiny\it D}_i}(z_i, \vec{r}_i)|^2 A_i
\label{average}
\ee
and the dipole orientation $\phi_i$ is defined as the angle between
transverse momentum $\vec{k}_{\!\perp}$ and dipole vector $\vec{r}_i$.
With~(\ref{dip_wf}) the integration over the dipole orientations
$\phi_i$ leads respectively to the Bessel function
$J_0(|\vec{k}_{\!\perp}||\vec{r}_{\!\mbox{\tiny\it D}_i}|)$ and the
generalized hypergeometric function\footnote{A review of generalized
  hypergeometric functions can be found in~\cite{MOS}. In the computer
  program Mathematica~\cite{Math} ``HypergeometricPFQ$[\{ {-1/2}\},\{
  {1/2,1}\}, -k^2_{\!\perp} r^2_{\!\mbox{\tiny\it D}}/4]$'' denotes
  this function.} $_1F_2 (-1/2;1/2,1;-k_{\!\perp}^2
r_{\!\mbox{\tiny\it D}_i}^2/4)$
\bea
\!\!\!\!\!\!\!\!\!\!\!
    \langle \psi_{\!\mbox{\tiny\it
    D}_i}|1-e^{i\vec{k}_{\!\perp}\vec{r}_i}|\psi_{\!\mbox{\tiny\it
    D}_i}\rangle 
&=& 
       \frac{1}{2\pi} \int_0^{2\pi} d\phi_{i}
       (1-e^{i\vec{k}_{\!\perp}
       \vec{r}_{\!\mbox{\tiny\it D}_i}})
\nonumber \\
&=& 
        1- J_0(|\vec{k}_{\!\perp}||\vec{r}_{\!\mbox{\tiny\it D}_i}|)
          \ ,
\label{average_dip}\\
\vspace{1cm}
\nonumber\\
\!\!\!\!\!\!\!\!\!\!\!
     \langle \psi_{\!\mbox{\tiny\it D}_i}|\tan^2\!\!\phi_i 
     (1-e^{i\vec{k}_{\!\perp}\vec{r}_i}) 
     |\psi_{\!\mbox{\tiny\it D}_i}\rangle &=& 
            \frac{1}{2\pi} \int_0^{2\pi} d\phi_{i}
            \tan^2\!\!\phi_{i} (1-e^{i\vec{k}_{\!\perp}
            \vec{r}_{\!\mbox{\tiny\it D}_i}}) \nonumber \\
            &=& -1+\ \!\! _1F_2 (-\frac{1}{2};\frac{1}{2},1; 
            \frac{-k_{\!\perp}^2 r_{\!\mbox{\tiny\it D}_i}^2}{4}) \  
        \ .
\label{average_string}
\eea
The important implications will be discussed in the next section.

\begin{figure}[h!]
\setlength{\unitlength}{1.cm}
\begin{center}
\epsfig{file=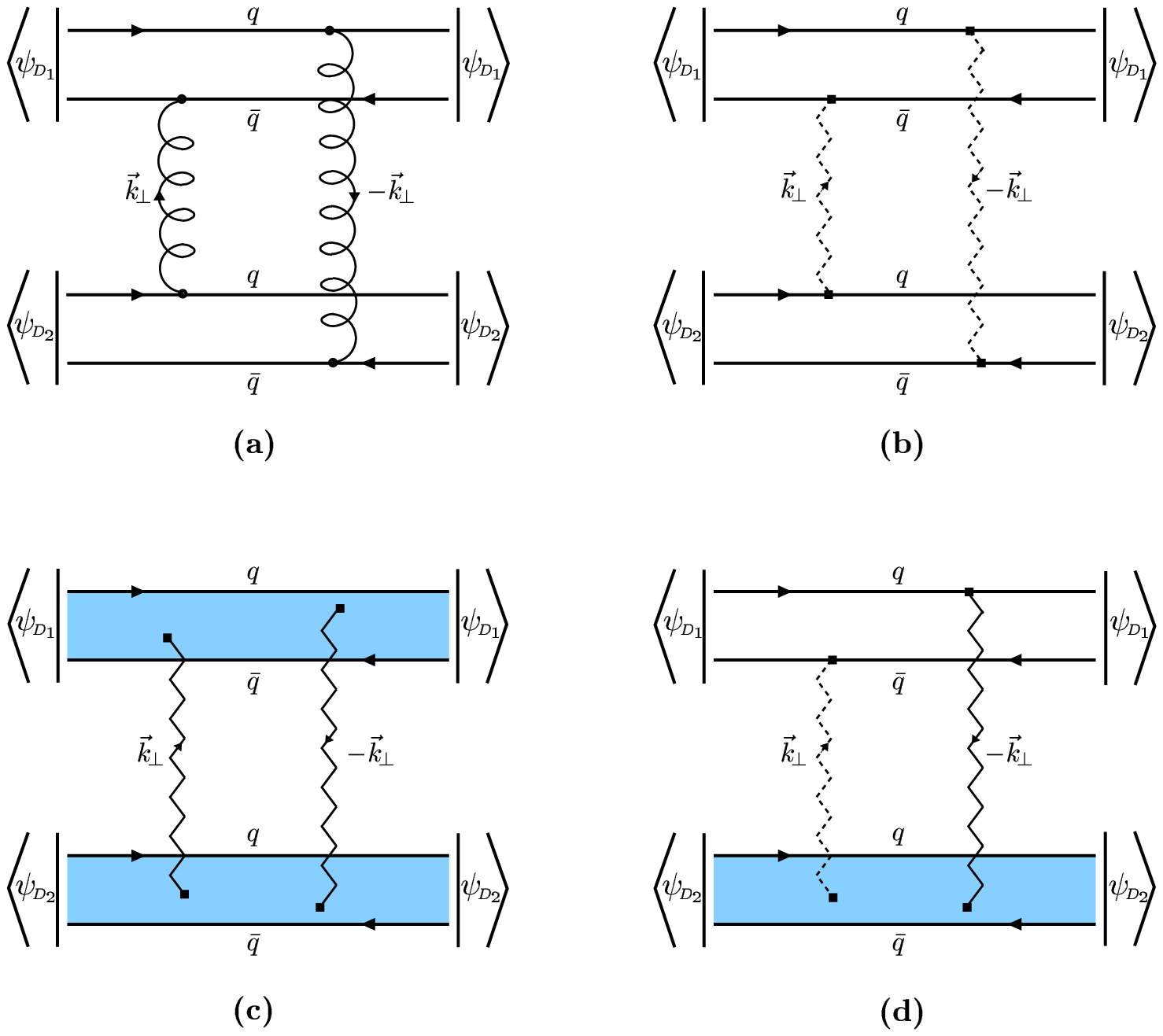, width=14.5cm}
\end{center}
\caption{\small Perturbative and non-perturbative contributions to dipole-dipole scattering:
  ${\bf (a)}$ perturbative quark-quark interaction and
  non-perturbative ${\bf (b)}$ quark-quark, ${\bf (c)}$ string-string,
  and ${\bf (d)}$ quark-string interactions. The term quark is used
  genuinely for quarks and antiquarks. Only one diagram is shown for
  each type of interaction. Paired horizontal lines represent
  quark-antiquark dipoles, surrounding brackets indicate the averages
  over the dipole orientations~(\ref{average}), shaded areas visualize
  strings, curly lines denote exchanged perturbative gluons, and
  dashed and solid zig-zag lines symbolize respectively the
  non-perturbative non-confining and confining correlation functions.}
\label{Fig_dipole_dipole_interactions}
\end{figure}

The integrand $I^{\pert}$ given in~(\ref{Ip}) describes the known
perturbative two-gluon\footnote{The exact $T$-matrix
  element~(\ref{Eq_model_purely_imaginary_T_amplitude_almost_final_result})
  important to describe scattering processes at ultra-high
  energies~\cite{Shoshi:2002in} goes beyond two-gluon exchange due to
  the higher orders in the cosine expansion.}
exchange~\cite{Low:1975sv+X,Gunion:iy} between the quarks and
antiquarks of the two dipoles. The ingredients of $I^{\pert}$ are
visualized for one combination of gluon exchanges in
Fig.~\ref{Fig_dipole_dipole_interactions}a: the paired horizontal lines
represent the {\em dipole factors}
$(1-e^{i\vec{k}_{\!\perp}\vec{r_i}})$ that describe the phase
difference between the quark and antiquark at separate transverse
positions, the surrounding brackets indicate the average over the
dipole orientations~(\ref{average}), the two curly lines illustrate
the gluon propagator squared $[i\tilde{D}_{\pert}^{\prime \,(2)}]^2$, and the
four vertices (dots) correspond to the strong coupling to the fourth
power $g^4=(4\pi\alphaS)^2$.

The integrand $I^{\nprt} = I^{\nprt}_{qq} + I^{\nprt}_{ss}$ given
in~(\ref{Eq_I_NP_qq}) and~(\ref{Eq_I_NP_ss}) describes the
non-perturbative interactions: the quark-quark, string-string, and
quark-string interactions identified by the appropriate correlation
functions $[i\tilde{D}^{\prime \,(2)}_1]^2$, $[i\tilde{D}^{(2)}]^2$, and $[i\tilde{D}^{\prime
  \,(2)}_1\,i\tilde{D}^{(2)}]$. These interactions are illustrated in
Figs.~\ref{Fig_dipole_dipole_interactions}b, \ref{Fig_dipole_dipole_interactions}c and
\ref{Fig_dipole_dipole_interactions}d, respectively. Analogous to the
perturbative interaction in Fig.~\ref{Fig_dipole_dipole_interactions}a, the
dashed and solid zig-zag lines represent respectively the
non-confining $(1-\kappa)\,i\tilde{D}^{\prime \,(2)}_1$ and the confining
$(\kappa/k^2_{\!\perp})\,i\tilde{D}^{(2)}$ non-perturbative correlations, the
shaded areas symbolize the strings, and the four vertices (squares) in
each figure indicate the ''non-perturbative coupling '' to the fourth
power $g^4_{\nprt} := \left(\pi^2G_2/24\right)^2$.

The integrand $I^{\nprt}_{qq}$ describes the non-perturbative {\em
  interactions between the quarks and antiquarks} of the two dipoles
and exhibits the same dipole factors
$(1-e^{i\vec{k}_{\!\perp}\vec{r_i}})$ that appear in the perturbative
integrand~(\ref{Ip}). $I^{\nprt}_{qq}$ contains three components: the
non-confining component, the confining component, and their
interference term. While the non-confining component visualized in
Fig.~\ref{Fig_dipole_dipole_interactions}b has the same structure as the
perturbative contribution, see also~(\ref{chi_p}) and~(\ref{chi_nc}),
the confining component shown in its more general form in
Fig.~\ref{Fig_dipole_dipole_interactions}c comes from the interaction between
the quarks and antiquarks at the endpoints of the strings, which will
be further discussed below Eq.~(\ref{ang_dep}). The interference term
describes the quark-string interaction as illustrated in
Fig.~\ref{Fig_dipole_dipole_interactions}d. Note that it reduces entirely to an
interaction between the quarks and antiquarks of the dipoles with the
additional denominator $1/k^2_{\!\perp}$ generated by the integrations
over the variables $v_1$ and $v_2$ in the confining component
$\chi^{\nprt}_c$~(\ref{chi_c}).

The integrand $I^{\nprt}_{ss}$ describes the non-perturbative {\em
  string-string interaction} shown in
Fig.~\ref{Fig_dipole_dipole_interactions}c.  The new angular dependencies in
the string-string interaction, the modified dipole factors
$\tan^2\!\!\phi_{i} (1-e^{i\vec{k}_{\!\perp} \vec{r}_{\!\mbox{\tiny\it
      D}_i}})$ in~(\ref{Eq_I_NP_ss}), are obtained as follows. The
integrations over $v_1$, $v_2$, $v_1^{\prime}$, and $v_2^{\prime}$ in
$(\chi^{\nprt}_c)^2$ produce the dipole factors
$(1-e^{i\vec{k}_{\!\perp}\vec{r_i}})$ and the denominator
$1/((\vec{k}_{\!\perp}\vec{r}_1)^2 (\vec{k}_{\!\perp}\vec{r}_2)^2)$.
This denominator multiplied with the additional factor
$(\vec{r_1}\vec{r_2})^2$ from $(\chi_c^{\nprt})^2$, see~(\ref{chi_c}),
gives the total angular dependence
\bea
\frac{(\vec{r}_1\vec{r}_2)^2}
{(\vec{k}_{\!\perp}\vec{r}_1)^2
(\vec{k}_{\!\perp}\vec{r}_2)^2} 
&=& 
   \frac{r^2_1r^2_2\cos^2(\phi_1-\phi_2)}
   {\left(k^2_{\!\perp}r^2_1\cos^2\!\phi_1\right)
   \left(k^2_{\!\perp}r^2_2\cos^2\!\phi_2\right)} \nonumber\\
&=& 
   \frac{\left(\cos\phi_1\cos\phi_2+\sin\phi_1\sin\phi_2\right)^2}
   {k^4_{\!\perp}\cos^2\!\phi_1\cos^2\!\phi_2} \nonumber\\
&=&
   \frac{1}{k^4_{\!\perp}} \left(1+2\tan\!\phi_1\tan\!\phi_2+
   \tan^2\!\phi_1\tan^2\!\phi_2 \right) \ .
\label{ang_dep}
\eea
The first term in~(\ref{ang_dep}) explains the interaction between the
endpoints of the strings with the additional factor $1/k^4_{\!\perp}$
in the first contribution $I^{\nprt}_{qq}$ already mentioned above.
The product of the second term in ~(\ref{ang_dep}),
$2\tan\!\phi_1\tan\!\phi_2$, with the dipole factors vanishes after
the integration over the dipole orientations $\phi_1$ and $\phi_2$.
The third term in~(\ref{ang_dep}), $\tan^2\!\phi_1\tan^2\!\phi_2$,
weights the different orientations of the strings and is
characteristic for the string-string interaction~(\ref{Eq_I_NP_ss}) in
our model. Due to this factor, the string-string interaction differs
significantly from the interaction between the quarks and antiquarks
of the dipoles known from perturbative two-gluon exchange~(\ref{Ip}).

\begin{figure}[htp]
\setlength{\unitlength}{1.cm}
\begin{center}
\epsfig{file=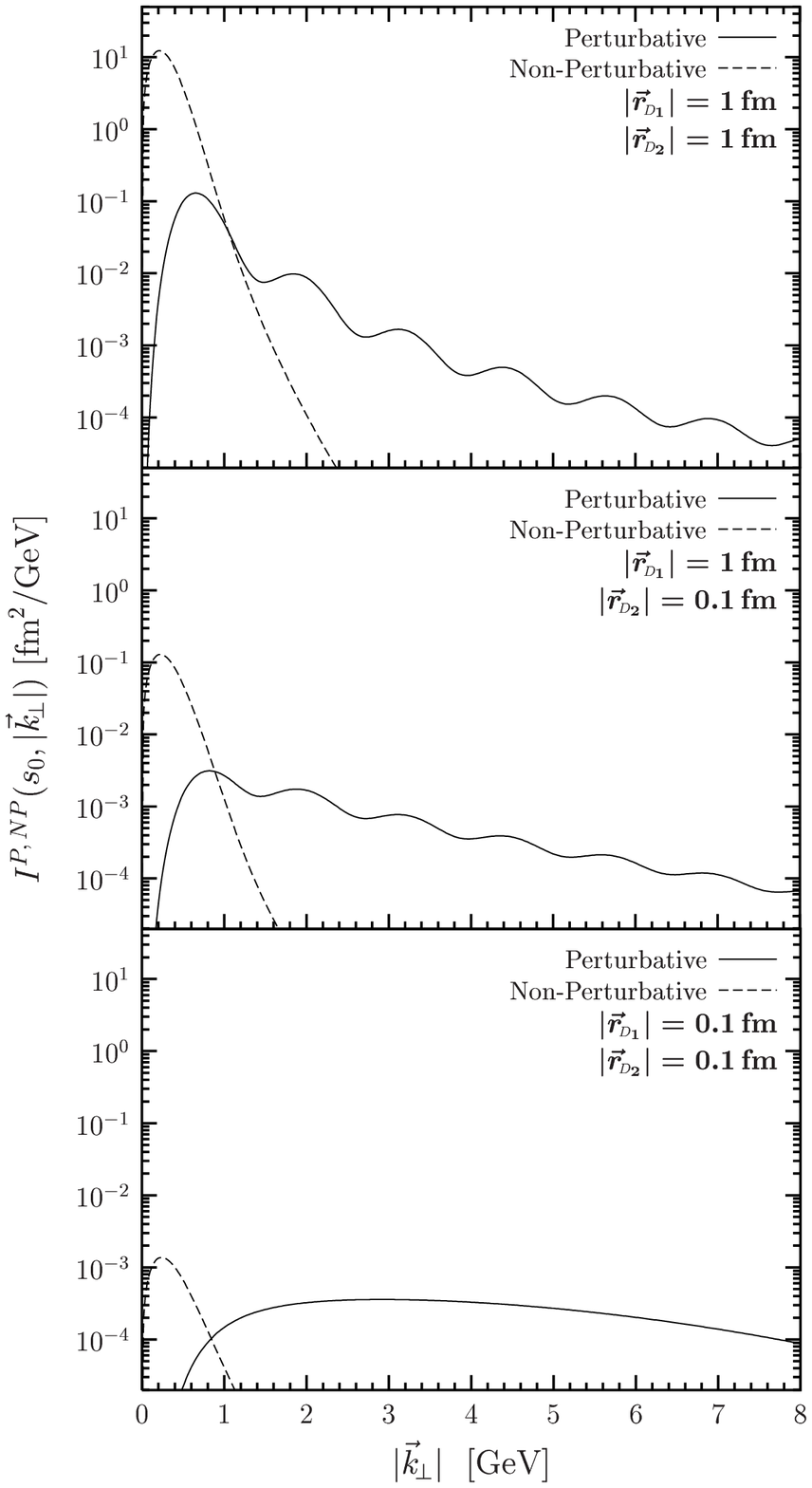, width=9.cm}
\end{center}
\caption{\small The perturbative 
  integrand $I^{\pert}$ (solid line) and the non-perturbative
  integrand $I^{\nprt}$ (dashed line) of the total dipole-dipole cross
  section~(\ref{sigt}) as a function of transverse momentum
  $|\vec{k}_{\!\perp}|$ for various dipole sizes
  $|\vec{r}_{\!\mbox{\tiny\it D}_1}|$ and $|\vec{r}_{\!\mbox{\tiny\it
      D}_2}|$.  The integrands are shown for the parameters given in
  Sec.~\ref{Sec_The_Loop_Loop_Correlation_Model} that allow a good description of the
  experimental data of hadron and photon reactions at $\sqrt{s_0}
  \approx 20\,\GeV$ with the wave functions given in
  Appendix~\ref{Sec_Wave_Functions} and the exact $T$-matrix
  element~(\ref{Eq_model_purely_imaginary_T_amplitude_almost_final_result}).
  The oscillations of the perturbative integrand $I^{\pert}$ originate
  from the Bessel function
  $J_0(|\vec{k}_{\!\perp}||\vec{r}_{\!\mbox{\tiny\it D}_i}|)$.  }
\label{integrand}
\end{figure}
In Fig.~\ref{integrand}, we show the perturbative integrand
$I^{\pert}$ (solid line) and the non-perturbative integrand
$I^{\nprt}$ (dashed line) of the total dipole-dipole cross
section~(\ref{sigt}) as a function of transverse momentum
$|\vec{k}_{\!\perp}|$ for various dipole sizes. The integrands have
been calculated with the model parameters given in
Sec.~\ref{Sec_The_Loop_Loop_Correlation_Model} that allow a good
description of the experimental data of hadron and photon reactions at
the c.m.\ energy of $\sqrt{s_0} \approx 20\,\GeV$ with the wave
functions given in Appendix~\ref{Sec_Wave_Functions} and the exact
$T$-matrix
element~(\ref{Eq_model_purely_imaginary_T_amplitude_almost_final_result}).
Evidently, the non-perturbative integrand $I^{\nprt}$ governs the low
momenta and the perturbative integrand $I^{\pert}$ the high momenta
exchanged in the dipole-dipole scattering. This behavior is a direct
consequence of the correlation functions: The non-perturbative
correlation functions~(\ref{Eq_D_(prime)(k^2)_for_exp_correlation})
and~(\ref{Eq_D(k^2)_for_exp_correlation}) favor low momenta and
suppress high momenta in comparison to the perturbative correlation
function~(\ref{Eq_massive_D_pge_prime}). For fixed dipole sizes
$|\vec{r}_{\!\mbox{\tiny\it D}_1}|$ and $|\vec{r}_{\!\mbox{\tiny\it
    D}_2}|$ the absolute values of the perturbative and
non-perturbative integrand are controlled respectively by the
parameters ($m_g$, $M$) and ($G_2$, $a$, $\kappa$) given in
Sec.~\ref{Sec_The_Loop_Loop_Correlation_Model}. With decreasing dipole
sizes the ratio $I^{\pert}/I^{\nprt}$ increases: for large dipole
sizes $|\vec{r}_{\!\mbox{\tiny\it D}_1}| = |\vec{r}_{\!\mbox{\tiny\it
    D}_2}| = 1\, \fm$, the non-perturbative contribution gives the
main contribution to the total dipole-dipole cross section; for small
dipole sizes $|\vec{r}_{\!\mbox{\tiny\it D}_1}| =
|\vec{r}_{\!\mbox{\tiny\it D}_2}| = 0.1\, \fm$, the perturbative
contribution dominates.

As can be seen from the analytic
results~(\ref{Ip})--(\ref{average_string}), the total dipole-dipole
cross section~(\ref{sigt}) or the forward scattering amplitude
$T(s_0,t=0)$ does not depend on the longitudinal quark momentum
fractions $z_{\!\mbox{\tiny\it D}_i}$ of the dipoles. For $t=0$ the
parameter $z_{\!\mbox{\tiny\it D}_i}$ disappears upon the integration
over $z_i$ since only $|\psi_{\!\mbox{\tiny\it
    D}_i}(z_i,\vec{r}_i)|^2$~(\ref{dip_wf}) depends on $z_i$.

The structure presented for dipole-dipole scattering remains in
reactions involving hadrons and photons: the hadronic and photonic
total cross sections are obtained from the total dipole-dipole cross
section~(\ref{sigtot_DD}) by replacing $|\psi_{\!\mbox{\tiny\it
    D}_i}(z_i,\vec{r}_i)|^2$ given in~(\ref{dip_wf}) with the hadron
and photon wave functions given in Appendix~\ref{Sec_Wave_Functions}.
As the total dipole-dipole cross section, the total hadronic and
photonic cross sections are independent of the parameters which
control the $z_i$\,-\,distribution in the wave functions due to the
normalization of the $z_i$\,-\,distributions.  The independence of the
total hadronic cross section on the widths $\Delta z_h$ can be seen
immediately with the Gaussian hadron wave
functions~(\ref{Eq_hadron_wave_function})
\bea
\!\!\!\!\!\!\!\!\!\!\!
\langle \psi_{h}|1-e^{i\vec{k}_{\!\perp}\vec{r}_i}|\psi_{h}\rangle &=& 
      1-e^{-\frac{1}{2} k_{\!\perp}^2 S_h^2} \ ,
\label{averages_explicit_qq}\\
\!\!\!\!\!\!\!\!\!\!\!
      \langle \psi_{h}|\tan^2\!\!\phi_i (1-e^{i\vec{k}_{\!\perp}\vec{r}_i}) 
      |\psi_{h}\rangle &=& 
      -1+e^{-\frac{1}{2} k_{\!\perp}^2 S_h^2}
      + \sqrt{\frac{\pi}{2}} \,|\vec{k}_{\!\perp}| S_h\, 
      \mbox{Erf}\left (\frac{|\vec{k}_{\!\perp}| S_h}{\sqrt{2}}\right)
\label{averages_explicit_ss}
\eea
with the error function $\mbox{Erf}(z)=\sqrt{2/\pi}\int_0^z dt
\exp(-t^2)$. These analytical results confirm the $\Delta z_h$ and
$z_{\!\mbox{\tiny\it D}}$\,-\,independence of the total dipole-proton
cross section assumed in phenomenological
models~\cite{Forshaw:1999uf,Golec-Biernat:1998js}.  However, the
non-forward hadronic scattering amplitude $T(s_0, t\neq 0)$ depends on
the parameter $\Delta z_h$ as shown explicitly in
Appendix~\ref{App_T_tneq0}.  Thus, the differential elastic cross
section $d\sigma^{el}/dt(s,t)$~(\ref{Eq_dsigma_el_dt}) and its
logarithmic slope $B(s,t)$~(\ref{Eq_elastic_local_slope}) are $\Delta
z_h$\,-\,dependent. In fact, this $\Delta z_h$\,-\,dependence is
essential for the agreement with experimental
data~\cite{Shoshi:2002in}.

The $|\vec{k}_{\!\perp}|$\,-\,dependence of the perturbative and
non-perturbative integrand for hadron-hadron, hadron-photon, and
photon-photon cross sections at high photon virtualities is similar to
the one of the perturbative and non-perturbative integrand of the
dipole-dipole cross section shown in Fig.~\ref{integrand} for
($|\vec{r}_{\!\mbox{\tiny\it D}_1}| = |\vec{r}_{\!\mbox{\tiny\it
    D}_2}| = 1\, \fm$), ($|\vec{r}_{\!\mbox{\tiny\it D}_1}| = 1\,
\fm$, $|\vec{r}_{\!\mbox{\tiny\it D}_2}| = 0.1\, \fm$), and
($|\vec{r}_{\!\mbox{\tiny\it D}_1}| = |\vec{r}_{\!\mbox{\tiny\it
    D}_2}| = 0.1\, \fm$), respectively. Of course, the absolute values
differ and the oscillations of the perturbative integrand caused by
the Bessel functions disappear.

\section{Decomposition of the QCD String into Dipoles and the Unintegrated Gluon Distribution}
\label{Sec_The_Decomposition_of_the_String_into_Dipoles_and_the_Unintegrated_Gluon_Distribution}

\bigskip

The {\em unintegrated gluon distribution} of hadrons\footnote{The word
  hadron and the subscript $h$ is used genuinely for hadrons and
  photons in this section.} ${\cal F}_{h}(x,k_{\!\perp}^2)$ is a
basic, universal quantity convenient for the computation of many
scattering observables at small $x$. It is the central object in the
BFKL~\cite{Kuraev:fs+X} and CCFM~\cite{Ciafaloni:1987ur+X} evolution
equations. Upon integration over the transverse gluon momentum
$|\vec{k}_{\!\perp}|$ it leads to the conventional gluon distribution
$xG_{h}(x,Q^2)$ used in the DGLAP evolution
equation~\cite{Gribov:ri+X}. The unintegrated gluon distribution is
crucial to describe processes in which transverse momenta are
explicitly exposed such as dijet~\cite{Nikolaev:1994cd+X} or vector
meson~\cite{Nemchik:1997xb} production at HERA. Its explicit
$|\vec{k}_{\!\perp}|$ dependence is particularly suited to study the
interplay between soft and hard physics. In this section, an exact
representation of the string as a collection of stringless
quark-antiquark dipoles is presented that allows to extract the
perturbative and non-perturbative contributions to ${\cal
  F}_{h}(x,k_{\!\perp}^2)$ from our total dipole-hadron cross section
via $|\vec{k}_{\!\perp}|$\,-\,factorization.

\medskip

We calculate the unintegrated gluon distribution ${\cal
  F}_{h}(x,k_{\!\perp}^2)$ with the $T$-matrix element in the limit of
small
${\chi}$-functions~(\ref{Eq_model_purely_imaginary_T_amplitude_small_chi_limit}).
Motivated by the successful description of many data in our recent
work~\cite{Shoshi:2002in} we give a strong energy dependence to the
perturbative contribution ${\chi^{\pert}}$ and a weak one to the
non-perturbative contribution ${\chi^{\nprt}}$,
\bea 
        ({\chi^{\pert}})^2 & \to & 
        ({\chi^{\pert}})^2 \left( \frac{x_0}{x}\right)^{\epsilon^{\pert}}
\nonumber \\
        ({\chi^{\nprt}})^2 & \to & ({\chi^{\nprt}})^2
        \left(\frac{x_0}{x}\right)^{\epsilon^{\nprt}} 
\label{x_dep}
\eea
where the values of the exponents ${\epsilon^{\pert}} = 0.73$ and
${\epsilon^{\nprt}} = 0.125$ are adopted from~\cite{Shoshi:2002in} and
$x_0 = 2.4\cdot10^{-3}$ is adjusted to reproduce at $Q^2=1\,\GeV^2$
the integrated gluon distribution of the proton $xG_{\!p}(x,Q^2)$
extracted from the HERA data~\cite{Abramowicz:1998ii+X}.
The small-${\chi}$
limit~(\ref{Eq_model_purely_imaginary_T_amplitude_small_chi_limit})
considered here is applicable only for $x = Q^2/s \ge 10^{-4}$. At
higher energies (lower Bjorken\,-\,$x$) the full $T$-matrix
element~(\ref{Eq_model_purely_imaginary_T_amplitude_almost_final_result})
has to be used in order to satisfy unitarity constraints of the
$S$-matrix~\cite{Shoshi:2002in,Shoshi:2002ri}.

\medskip

The total dipole-hadron cross section
$\sigma_{\!Dh}(x,|\vec{r}_{\!\mbox{\tiny\it D}}|)$ is obtained from
the total dipole-dipole cross section~(\ref{sigtot_DD}) by replacing
$|\psi_{\!\mbox{\tiny\it D}_2}(z_2,\vec{r}_2)|^2$ with a squared
hadron wave function $|\psi_{\!\mbox{\tiny\it h}}(z_2,\vec{r}_2)|^2$.
Accordingly, the $x$-dependent total dipole-hadron cross section reads
\bea
&&\!\!\!\!\!\!\!\!\!\!
\sigma_{\!Dh}(x, |\vec{r}_{\!\mbox{\tiny\it D}}|) = 
    \frac{8}{9}\frac{1}{4\pi} \int dk_{\!\perp}^2 
\label{sigdip}\\
    &&\!\!\!\!\!\!\!\!\!\!\times
    \left [
    \left (4\pi \alphaS(k^2_{\!\perp})\right)^2
    \left \{ \left [i\tilde{D}_{\pert}^{\prime \,(2)}(k^2_{\!\perp})\right ]^2
    \left(1-J_0(|\vec{k}_{\!\perp}||\vec{r}_{\!\mbox{\tiny\it D}}|)
    \right) 
    \langle \psi_{h}|1-e^{i\vec{k}_{\!\perp}\vec{r}_2}|\psi_{h}\rangle
    \right \} \left ( \frac{x_0}{x} \right )^{\!\!\epsilon^{\pert}} \right .
    \nonumber \\
    &&\!\!\!\!\!\!\!\!\!\!\!
    + \!\left(\!\frac{\pi^2G_2}{24}\!\right)^{\!\!2}\!
    \left \{ \!\!
    \left[\frac{\kappa}{k_{\!\perp}^2}\, 
    i\tilde{D}^{(2)}(k_{\!\perp}^2)
    + (1 - \kappa)\,
    i\tilde{D}^{\prime \,(2)}(k_{\!\perp}^2)\! 
    \right]^2\!\!
    \left(1-J_0(|\vec{k}_{\!\perp}||\vec{r}_{\!\mbox{\tiny\it D}}|)
    \right) 
    \langle \psi_{h}|1-e^{i\vec{k}_{\!\perp}\vec{r}_2}|\psi_{h}\rangle
    \right .
    \nonumber \\
    &&\!\!\!\!\!\!\!\!\!\!\!\!
    \left . \left . 
    +\! 
    \left[\frac{\kappa}{k_{\!\perp}^2}\,i\tilde{D}^{(2)}(k_{\!\perp}^2)\right]^2 
    \!\!\left(-1\!+\ \!\! _1F_2 (-\frac{1}{2};\frac{1}{2},1; 
    \frac{-k_{\!\perp}^2 r_{\!\mbox{\tiny\it D}}^2}{4})\right)
    \!\langle \psi_{h}|\tan^2\!\!\phi_2(1-e^{i\vec{k}_{\!\perp}\vec{r}_2}) 
    |\psi_{h}\rangle\! 
    \right \}\!\left ( \frac{x_0}{x} \right )^{\!\!\epsilon^{\nprt}} \right
    ] \nonumber
\eea
with the Bessel function $J_0(|\vec{k}_{\!\perp}||\vec{r}_{\!\mbox{\tiny\it D}}|)$ and the
generalized hypergeometric function $_1F_2 (-1/2;1/2,1;-k_{\!\perp}^2
r_{\!\mbox{\tiny\it D}}^2/4)$ derived in the previous section.

For dipole sizes $|\vec{r}_{\!\mbox{\tiny\it D}}| \to 0$, the
perturbative contribution to $\sigma_{\!Dh}(x,
|\vec{r}_{\!\mbox{\tiny\it D}}|)$ is known to vanish quadratically
with decreasing dipole size,
$\sigma_{\!Dh}(x,|\vec{r}_{\!\mbox{\tiny\it D}}|) \propto
r^2_{\!\mbox{\tiny\it D}}$. This behavior reflects the weak absorption
of a small color-singlet dipole in the hadron and is known as {\em
  color transparency}. It can be seen immediately from~(\ref{sigdip}) as
\be
\left (1-J_0(|\vec{k}_{\!\perp}||\vec{r}_{\!\mbox{\tiny\it D}}|)\right) \approx 
\frac{k_{\!\perp}^2 r_{\!\mbox{\tiny\it D}}^2}{4} \quad \mbox{for
$|\vec{r}_{\!\mbox{\tiny\it D}}| \to 0$ and finite $|\vec{k}_{\!\perp}|$}. 
\label{bess_small_r}
\ee
The non-perturbative contribution to $\sigma_{\!Dh}(x,
|\vec{r}_{\!\mbox{\tiny\it D}}|)$ gives color transparency as well
since the generalized hypergeometric function behaves as
\be
\left(-1+\ \!\!_1F_2
(-\frac{1}{2};\frac{1}{2},1;\frac{-k_{\!\perp}^2 r_{\!\mbox{\tiny\it D}}^2}{4})\right) \approx 
\frac{k_{\!\perp}^2 r_{\!\mbox{\tiny\it D}}^2}{4} \quad \mbox{for
$|\vec{r}_{\!\mbox{\tiny\it D}}| \to 0$ and finite $|\vec{k}_{\!\perp}|$}.
\ee

For large dipole sizes, $|\vec{r}_{\!\mbox{\tiny\it D}}|\,\gtsim\,1\,
\fm$, the perturbative contribution to $\sigma_{\!Dh}(x,
|\vec{r}_{\!\mbox{\tiny\it D}}|)$ describing interactions of the quark
and antiquark of the dipole with the hadron saturates since
\be
\left (1-J_0(|\vec{k}_{\!\perp}||\vec{r}_{\!\mbox{\tiny\it D}}|)\right) \approx 
1 \quad \mbox{for
large $|\vec{k}_{\!\perp}||\vec{r}_{\!\mbox{\tiny\it D}}|$} \ .
\label{bess_large_r}
\ee
In contrast, the non-perturbative contribution to $\sigma_{\!Dh}(x,
|\vec{r}_{\!\mbox{\tiny\it D}}|)$ increases linearly with increasing
dipole size, $\sigma_{\!Dh}(x, |\vec{r}_{\!\mbox{\tiny\it D}}|)
\propto |\vec{r}_{\!\mbox{\tiny\it D}}|$.  This linear increase is
generated by the interaction of the string of the dipole with the
hadron: The string elongates linearly with the dipole size
$|\vec{r}_{\!\mbox{\tiny\it D}}|$ and, thus, has a linearly increasing
geometric cross section with the hadron. Indeed, this feature of the
string can be seen analytically since
\be
\left(-1+\ \!\!_1F_2
(-\frac{1}{2};\frac{1}{2},1;\frac{-k_{\!\perp}^2 r_{\!\mbox{\tiny\it D}}^2}{4})\right) \propto |\vec{k}_{\!\perp}||\vec{r}_{\!\mbox{\tiny\it D}}|
\quad \mbox{for
large $|\vec{k}_{\!\perp}||\vec{r}_{\!\mbox{\tiny\it D}}|$} \ .
\ee
When considered in Euclidean space-time, the same string gives also
the linear confining potential between a static quark and antiquark at
large $q{\bar q}$
separations~\cite{Rueter:1995cn,Euclidean_Model_Applications}.
Obviously, the behavior of the total dipole-hadron cross section is
related to the confining potential. Furthermore, as we are working in
the quenched approximation, there is no string breaking through
dynamical quark-antiquark production at large dipole sizes. String
breaking is expected to stop the linear increase of the total
dipole-hadron cross section at dipole sizes of
$|\vec{r}_{\!\mbox{\tiny\it D}}| \,\gtsim\, 1\, \fm$ analogous to the
saturation of the static $q{\bar q}$ potential seen for large $q{\bar
  q}$ separations on the lattice in full QCD~\cite{Laermann:1998gm,Bali:2000gf}.

\begin{figure}[htb]
\setlength{\unitlength}{1.cm}
\begin{center}
\epsfig{file=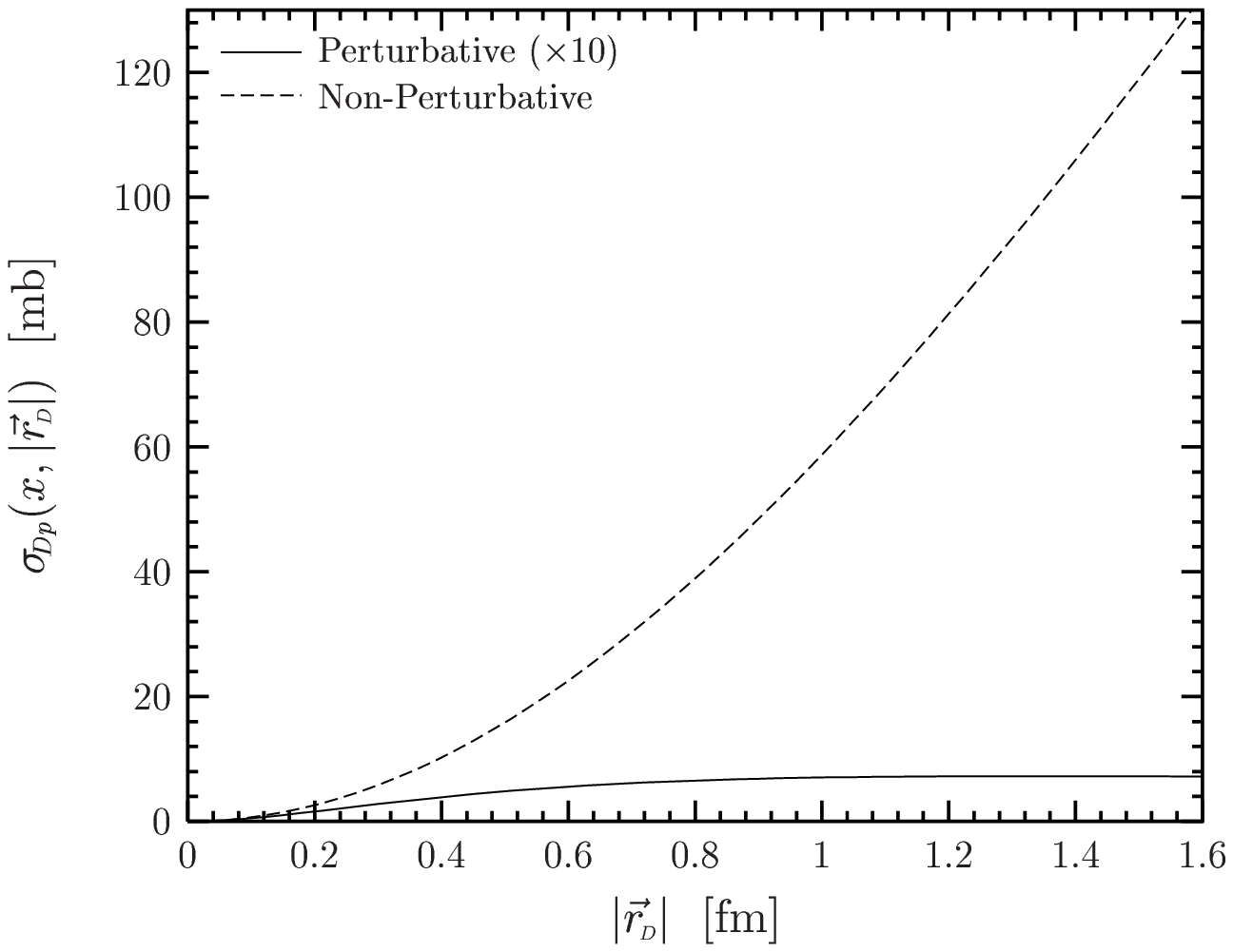,width=13.cm}
\end{center}
\caption{\small Perturbative (solid line) and
  non-perturbative (dashed line) contributions to the total
  dipole-proton cross section $\sigma_{\!Dp}(x,
  |\vec{r}_{\!\mbox{\tiny\it D}}|)$ as a function of the dipole size
  $|\vec{r}_{\!\mbox{\tiny\it D}}|$ at $x = x_0 = 2.4\cdot 10^{-3}$.
  with the perturbative contribution multiplied by a factor of $10$.
  The perturbative contribution shows color transparency at small
  $|\vec{r}_{\!\mbox{\tiny\it D}}|$ and saturates at large
  $|\vec{r}_{\!\mbox{\tiny\it D}}|$. The non-perturbative contribution
  shows also color transparency at small
  $|\vec{r}_{\!\mbox{\tiny\it D}}|$ but increases linearly with
  increasing $|\vec{r}_{\!\mbox{\tiny\it D}}|$.}
\label{sig_dipp}
\end{figure}

In Fig.~\ref{sig_dipp}, we show the perturbative (solid line) and
non-perturbative (dashed line) contributions to the total
dipole-proton cross section $\sigma_{\!Dp}(x,
|\vec{r}_{\!\mbox{\tiny\it D}}|)$ as a function of the dipole size
$|\vec{r}_{\!\mbox{\tiny\it D}}|$ at $x = x_0 = 2.4\cdot 10^{-3}$,
where the perturbative contribution is multiplied by a factor of $10$.
The dipole-proton cross section is computed with the simple Gaussian
proton wave function~(\ref{Eq_hadron_wave_function}) and illustrates
the general features discussed above: The perturbative contribution
shows color transparency at small $|\vec{r}_{\!\mbox{\tiny\it D}}|$
and saturates at large $|\vec{r}_{\!\mbox{\tiny\it D}}|$. The
non-perturbative contribution shows also color
transparency at small $|\vec{r}_{\!\mbox{\tiny\it D}}|$ but increases
linearly with increasing $|\vec{r}_{\!\mbox{\tiny\it D}}|$.  

Our result~(\ref{sigdip}) shows that the
$|\vec{k}_{\!\perp}|$\,-\,dependence of the hadron constituents
factorizes from the rest of the process in both the perturbative and
the non-perturbative contribution to the total dipole-hadron cross
section. This factorization -- known in perturbative QCD as
$|\vec{k}_{\!\perp}|$\,-\,factorization~\cite{Catani:1990xk+X} --
allows to define the unintegrated gluon distribution ${\cal
  F}_{h}(x,k^2_{\!\perp})$ as follows~\cite{Nikolaev:1991ja,Nikolaev:ce,Golec-Biernat:1999qd}
\be 
\sigma_{\!Dh}(x,|\vec{r}_{\!\mbox{\tiny\it D}}|) =
     \frac{4\pi^2r^2_{\!\mbox{\tiny\it D}}}{3}\int \!\!dk_{\!\perp}^2\,
     \frac{\left(1-J_0(|\vec{k}_{\!\perp}||\vec{r}_{\!\mbox{\tiny\it D}}|)
     \right)} {(|\vec{k}_{\!\perp}||\vec{r}_{\!\mbox{\tiny\it D}}|)^2}
     \,\alphaS(k^2_{\!\perp}) {\cal F}_{h}(x,k_{\!\perp}^2) \ .
\label{unint_gl_distr_def}
\ee 
For small dipole sizes
$|\vec{r}_{\!\mbox{\tiny\it D}}|$, this equation together
with~(\ref{bess_small_r}) and the integrated gluon
distribution
\be 
        xG_{h}(x,Q^2)=
        \int_0^{Q^2} \!dk_{\!\perp}^2 {\cal F}_{h}(x,k_{\!\perp}^2) 
\label{Eq_xG(x,Q^2)}   
\ee 
leads to the widely used perturbative QCD relation~\cite{Blaettel:rd}
\be
\sigma_{\!Dh}(x,|\vec{r}_{\!\mbox{\tiny\it D}}|) =
     \frac{\pi^2r^2_{\!\mbox{\tiny\it D}}}{3} 
     \left [ \alphaS(Q^2) xG_{h}(x,Q^2) \right ]_
     {\,Q^2 \,=\, c/r^2_{\!\mbox{\tiny\it D}}} \ ,
\label{rel_sigdip_xG}
\ee
where $c \approx 10$ is estimated from the properties of the Bessel
function $J_0(|\vec{k}_{\!\perp}||\vec{r}_{\!\mbox{\tiny\it
    D}}|)$~\cite{Nikolaev:ce+X}.

\begin{figure}[htb]
\setlength{\unitlength}{1.cm}
\begin{center}
\epsfig{file=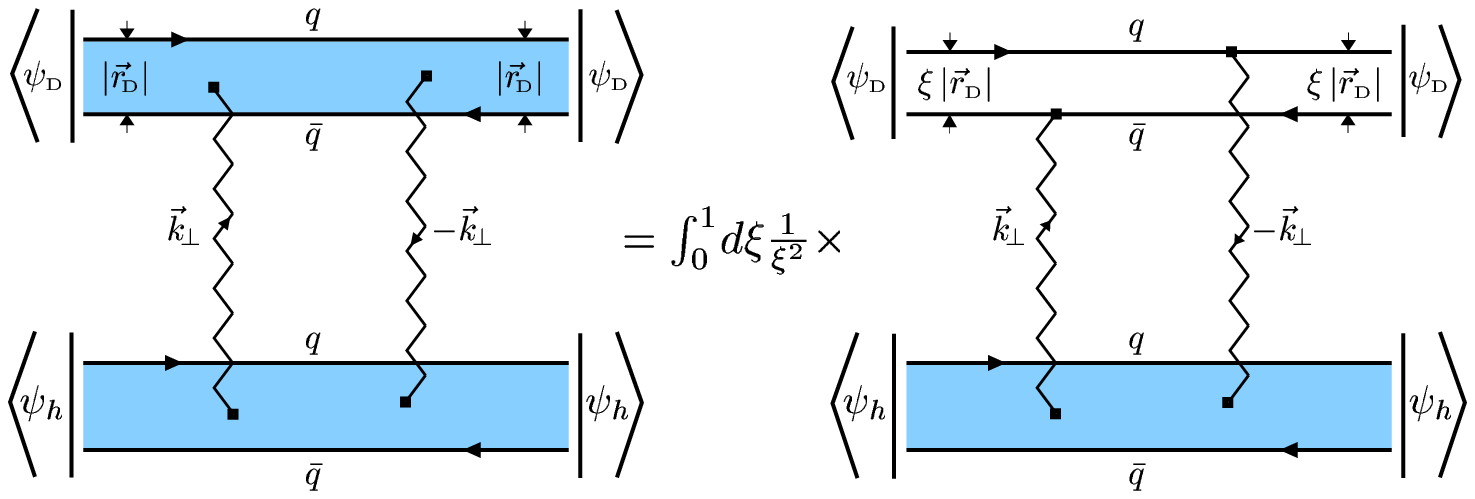,width=15.cm}
\end{center}
\caption{\small The string of length
  $|\vec{r}_{\!\mbox{\tiny\it D}}|$ is made up of stringless dipoles
  of size $\xi|\vec{r}_{\!\mbox{\tiny\it D}}|$ with $0\leq\xi\leq 1$
  and dipole number density $n(\xi) = 1/\xi^2$. The string-hadron
  scattering process reduces to an incoherent superposition of
  stringless dipole-hadron scattering processes.}
\label{sstoqs}
\end{figure}

To extract the unintegrated gluon distribution ${\cal
  F}_{h}(x,k^2_{\!\perp})$, we compare~(\ref{unint_gl_distr_def})
with (\ref{sigdip}) using the following mathematical identity
\be
\left(-1+\ \!\! _1F_2 (-\frac{1}{2};\frac{1}{2},1; 
\frac{-k_{\!\perp}^2 r_{\!\mbox{\tiny\it D}}^2}{4}) \right) = 
      \int_0^1\,d\xi \,\frac{1}{\xi^2}\left(1-J_0(|\vec{k}_{\!\perp}|
      |\vec{r}_{\!\mbox{\tiny\it D}}|\xi)\right) \ . 
\label{mathident}
\ee
As discussed in the previous section, the lhs of~(\ref{mathident})
results from the string averaged over all
orientations~(\ref{average_string}). Thus, the string confining the
quark-antiquark dipole of length $|\vec{r}_{\!\mbox{\tiny\it D}}|$ can
be represented as an integral over stringless dipoles of sizes
$\xi|\vec{r}_{\!\mbox{\tiny\it D}}|$ with $0\leq\xi\leq1$ and a dipole
number density of $n(\xi) = 1/\xi^2$.  As visualized in
Fig.~\ref{sstoqs}, the string-hadron scattering process reduces to an
incoherent superposition of stringless dipole-hadron scattering
processes with dipole sizes $0\leq \xi|\vec{r}_{\!\mbox{\tiny\it
    D}}|\leq|\vec{r}_{\!\mbox{\tiny\it D}}|$ and dipole number
density $n(\xi) = 1/\xi^2$. The decomposition of the string into many
smaller stringless dipoles via~(\ref{mathident}) behaves similar to
the wave function of a $q{\bar q}$ onium state in the large $N_c$
limit~\cite{Mueller:1994rr,Mueller:1994jq}: The numerous gluons
emitted inside the onium state can be considered as many $q{\bar q}$
dipoles in the large $N_c$ limit.

Inserting~(\ref{mathident}) into~(\ref{sigdip}) and rescaling the
momentum variable $|\vec{k}_{\!\perp}^{\prime}|=
\xi|\vec{k}_{\!\perp}|$, the string-hadron ($sh$) contribution to the
total dipole-hadron cross section~(\ref{sigdip}) becomes
\bea
\sigma_{\!Dh}^{\,sh}
(x, |\vec{r}_{\!\mbox{\tiny\it D}}|)&=& 
      \frac{8}{9}\frac{1}{4\pi} \int dk_{\!\perp}^{\prime \,2}
      \left(1-J_0(|\vec{k}_{\!\perp}^{\prime}|
      |\vec{r}_{\!\mbox{\tiny\it D}}|)\right)
\label{sigdip_sp} \\
      &&
      \hphantom{\hspace{-3cm}}
      \times 
      \left( \frac{\pi^2G_2}{24} \right)^2 \left \{ \frac{\kappa^2}
      {k_{\!\perp}^{\prime \,4}}
      \int_0^1 d\xi
      \left[i\tilde{D}^{(2)}(\frac{k_{\!\perp}^{\prime \,2}}{\xi^2})\right]^2 
      \langle \psi_{h}|\tan^2\!\!\phi_2
      (1-e^{i(\vec{k}_{\!\perp}^{\prime}/
      {\xi})\vec{r}_2})|\psi_{h} 
      \rangle \right \}\left ( \frac{x_0}{x} 
      \right )^{\!\!\epsilon^{\nprt}}. \nonumber
\eea
The dipole factor $(1-J_0(|\vec{k}_{\!\perp}^{\prime}|
|\vec{r}_{\!\mbox{\tiny\it D}}|)$ indicates that the string-hadron
interaction has been rewritten into a stringless dipole-hadron
interaction. The string confining the dipole has been shifted into the
hadron. Comparing (\ref{unint_gl_distr_def}) with our new expression for the total dipole-hadron cross section,
\bea
\!\!\!\!\!\!\!\!
\sigma_{\!Dh}(x, |\vec{r}_{\!\mbox{\tiny\it D}}|) &=& 
    \frac{8}{9}\frac{1}{4\pi} \int dk_{\!\perp}^2 
    \left(1-J_0(|\vec{k}_{\!\perp}||\vec{r}_{\!\mbox{\tiny\it D}}|)\right)
    \nonumber \\
    &&
    \hphantom{\hspace{-2cm}}
    \times
    \left [
    \left (4\pi\alphaS(k^2_{\!\perp})\right)^2
    \left \{ \left [i\tilde{D}_{\pert}^{\prime \,(2)}(k^2_{\!\perp})\right ]^2
    \langle \psi_{h}|1-e^{i\vec{k}_{\!\perp}\vec{r}_2}|\psi_{h} \rangle
    \right \} \left( \frac{x_0}{x} \right )^{\!\!\epsilon^{\pert}} \right .
    \nonumber \\
    &&
    \hphantom{\hspace{-2cm}}
    + \left(\frac{\pi^2G_2}{24}\right)^2
    \left \{ \!\!
    \left[(1-\kappa)
    \left[i\tilde{D}_1^{\prime \,(2)}(k_{\!\perp}^2)\right]+
    \frac{\kappa}{k_{\!\perp}^2} 
    \left[i\tilde{D}^{(2)}(k_{\!\perp}^2)\right]
    \right]^2
    \langle \psi_{h}|1-e^{i\vec{k}_{\!\perp}\vec{r}_2}|\psi_{h}\rangle
    \right .
    \nonumber \\
    &&
    \hphantom{\hspace{-2cm}}
    \left . \left .
    +\, \frac{\kappa^2}{k_{\!\perp}^4} \int_0^1 d\xi
    \left[i\tilde{D}^{(2)}(\frac{k_{\!\perp}^2}{\xi^2})\right]^2 
    \langle
    \psi_{h}|\tan^2\!\!\phi_2(1-e^{i(\vec{k}_{\!\perp}/{\xi})\vec{r}_2
    }) |\psi_{h}\rangle 
    \right \} \left( \frac{x_0}{x} \right )^{\!\!\epsilon^{\nprt}}
    \right ] \ ,
\label{sigdipxi}
\eea
one obtains the unintegrated gluon distribution 
\bea
\!\!\!\!\!
{\cal F}_{h}(x,k_{\!\perp}^2) &=& 
    \frac{k_{\!\perp}^2}{6\pi^3\alphaS(k^2_{\!\perp})}
    \nonumber \\
    &&
    \hphantom{\hspace{-2cm}}
    \times
    \left [
    \left (4\pi\alphaS(k^2_{\!\perp})\right)^2
    \left \{ \left [i\tilde{D}_{\pert}^{\prime \,(2)}(k^2_{\!\perp})\right ]^2
    \langle \psi_{h}|1-e^{i\vec{k}_{\!\perp}\vec{r}_2}|\psi_{h}\rangle
    \right \} \left( \frac{x_0}{x} \right )^{\!\!\epsilon^{\pert}} \right .
    \nonumber \\
    &&
    \hphantom{\hspace{-2cm}}
    +\left(\frac{\pi^2G_2}{24}\right)^2
    \left \{ \!\!
    \left[(1-\kappa)
    \left[i\tilde{D}_1^{\prime \,(2)}(k_{\!\perp}^2)\right] +
    \frac{\kappa}{k_{\!\perp}^2} 
    \left[i\tilde{D}^{(2)}(k_{\!\perp}^2)\right]
    \right]^2
    \langle \psi_{h}|1-e^{i\vec{k}_{\!\perp}\vec{r}_2}|\psi_{h}\rangle
    \right .
    \nonumber \\
    &&
    \hphantom{\hspace{-2cm}}
    \left . \left .
    + \frac{\kappa^2}{k_{\!\perp}^4}\int_0^1 d\xi
    \left[i\tilde{D}^{(2)}(\frac{k_{\!\perp}^2}{\xi^2})\right]^2 
    \langle
    \psi_{h}|\tan^2\!\!\phi_2(1-e^{i(\vec{k}_{\!\perp}/{\xi})
    \vec{r}_2})|\psi_{h}\rangle 
    \right \} \left( \frac{x_0}{x} \right )^{\!\!\epsilon^{\nprt}}
    \right ] \ .
\label{lf}
\eea
This result shows explicitly the microscopic structure of the
perturbative and non-perturbative contribution to ${\cal
  F}_{h}(x,k^2_{\!\perp})$. It is valid for any hadron wave
function. To get numerical values for the unintegrated gluon
distribution ${\cal F}_{h}(x,k^2_{\!\perp})$, the hadron wave
functions must be specified.

\section{Numerical Results for the Unintegrated Gluon Distribution in Hadrons and Photons}
\label{Sec_Numerical_Results_of_Unintegrated_Gluon_Distributions_of_Hadrons_and_Photons} 

In this section, we present the unintegrated gluon
distribution~(\ref{lf}) for protons, pions, kaons, and photons
computed with the Gaussian hadron wave
function~(\ref{Eq_hadron_wave_function}) and the perturbatively
derived photon wave
functions~(\ref{Eq_photon_wave_function_T_squared})
and~(\ref{Eq_photon_wave_function_L_squared}). To account for
non-perturbative effects at low photon virtuality $Q^2$ in the photon
wave functions, quark masses $m_f(Q^2)$ are used that interpolate
between the current quarks at large $Q^2$ and the constituent quarks
at small $Q^2$~\cite{Dosch:1998nw} as discussed in
Appendix~\ref{Sec_Wave_Functions}.

\begin{figure}[htb]
\setlength{\unitlength}{1.cm}
\begin{center}
\epsfig{file=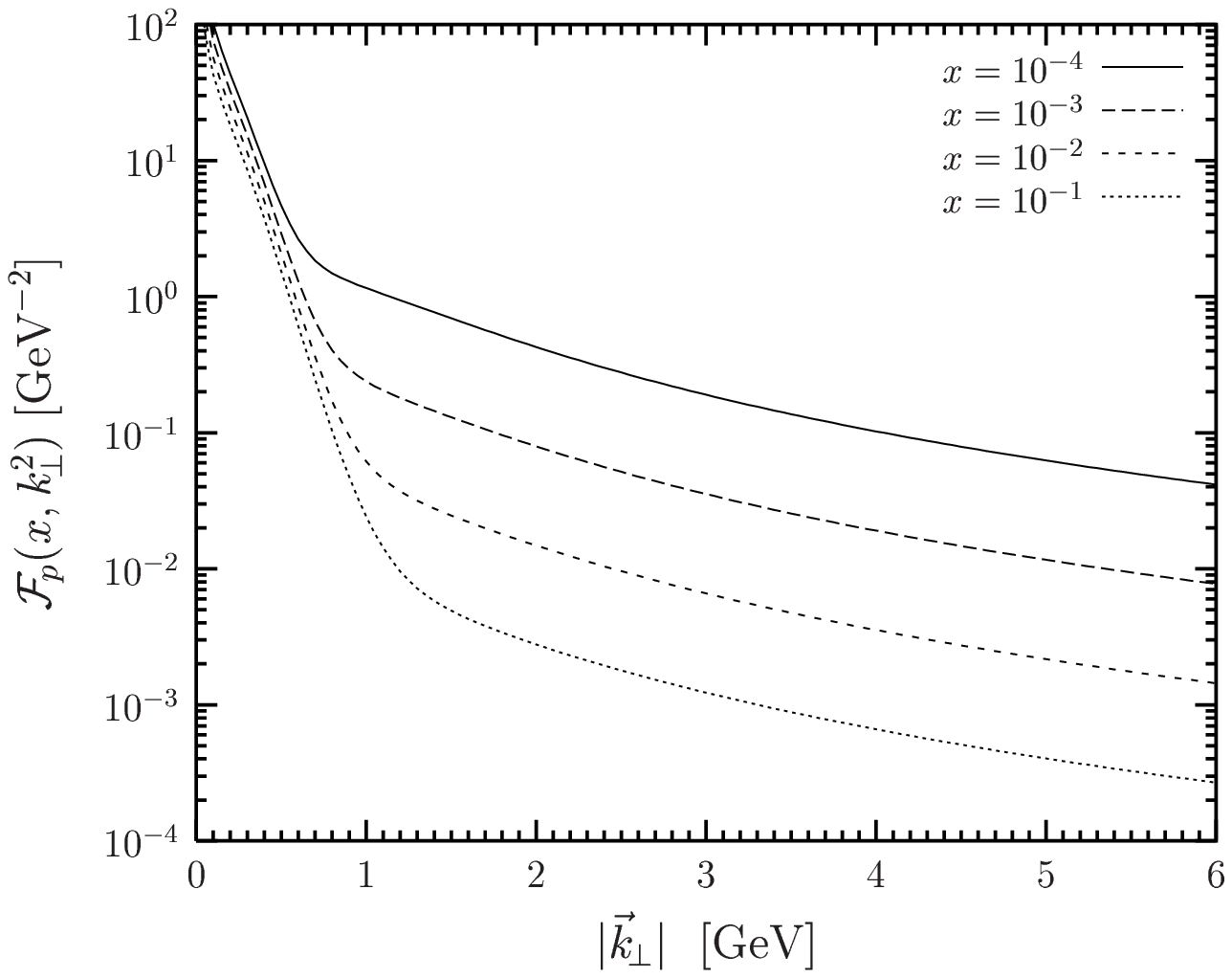,width=13.cm}
\end{center}
\caption{\small The unintegrated gluon distribution of the proton
  ${\cal F}_{\!p}(x,k_{\!\perp}^2)$ as a function of transverse
  momentum $|\vec{k}_{\!\perp}|$ at Bjorken\,-\,$x$ values of
  $10^{-1}$, $10^{-2}$, $10^{-3}$ and $10^{-4}$.}
\label{fg_k.eps}
\end{figure}

\begin{figure}[htb]
\setlength{\unitlength}{1.cm}
\begin{center}
\epsfig{file=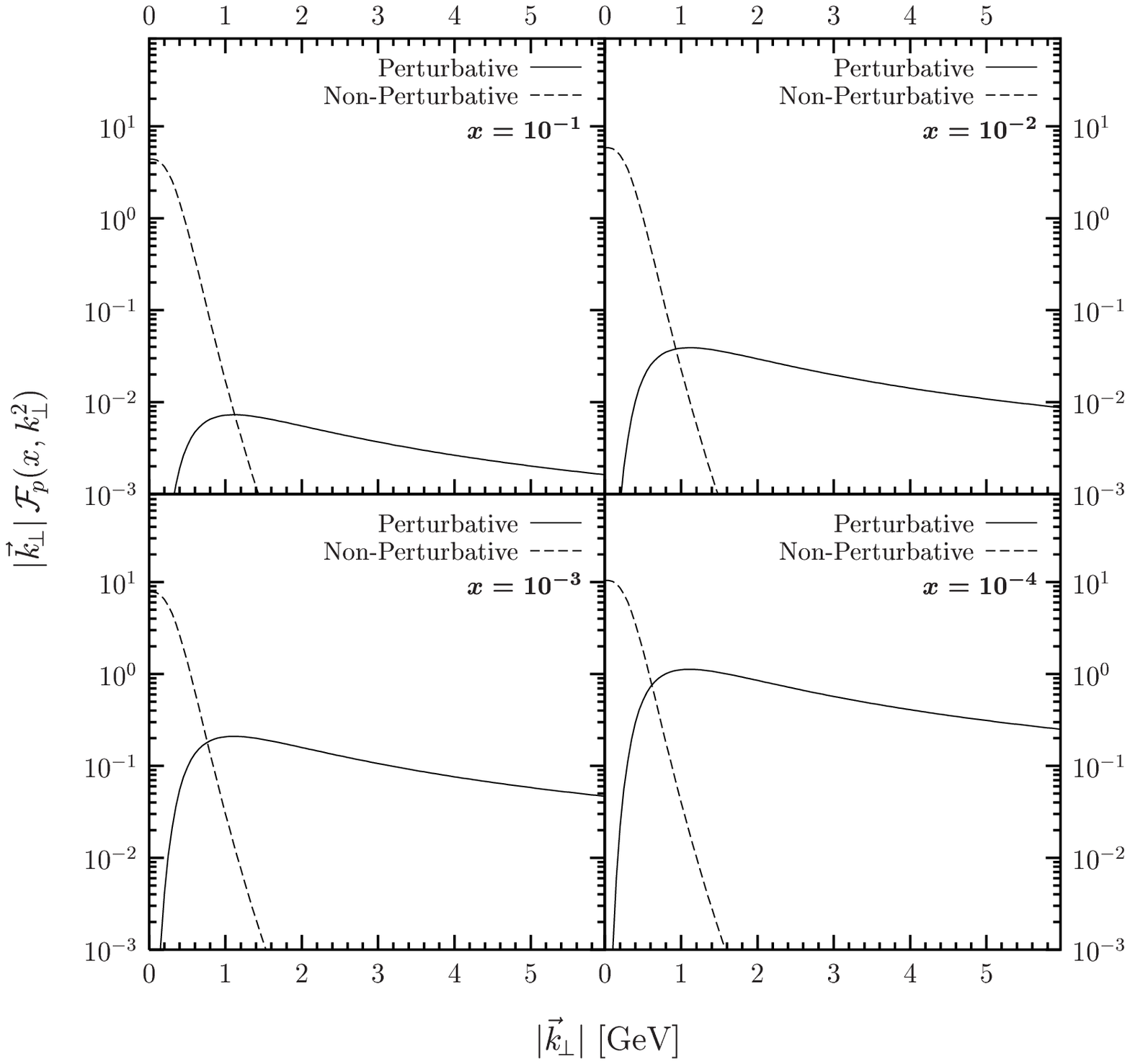,width=13cm}
\end{center}
\caption{\small The unintegrated gluon distribution of the proton ${\cal
    F}_{\!p}(x,k_{\!\perp}^2)$ times the tranverse momentum
  $|\vec{k}_{\!\perp}|$ as a function of $|\vec{k}_{\!\perp}|$ at
  Bjorken\,-\,$x$ values of $10^{-1}$, $10^{-2}$, $10^{-3}$ and
  $10^{-4}$.}
\label{p_fg_k_p_vs_np.eps}
\end{figure}

The unintegrated gluon distribution of the proton ${\cal
  F}_{\!p}(x,k_{\!\perp}^2)$ is shown as a function of transverse
momentum $|\vec{k}_{\!\perp}|$ at $x = 10^{-1}$, $10^{-2}$, $10^{-3}$,
and $10^{-4}$ in Fig.~\ref{fg_k.eps}. Figure~\ref{p_fg_k_p_vs_np.eps}
illustrates the interplay of the perturbative (solid line) and
non-perturbative (dashed line) contributions to
$|\vec{k}_{\!\perp}|{\cal F}_{\!p}(x,k_{\!\perp}^2)$ as a function of
transverse momentum $|\vec{k}_{\!\perp}|$ for the same values of $x$.
At small momenta $|\vec{k}_{\!\perp}|$, the unintegrated gluon
distribution is dominated by the non-perturbative contribution and
behaves as $1/|\vec{k}_{\!\perp}|$. This behavior reflects the linear
increase of the total dipole-proton cross section at large dipole
sizes. In contrast, the saturation model of Golec-Biernat and
W{\"u}sthoff~\cite{Golec-Biernat:1999qd} shows the behavior ${\cal
  F}^{\mbox{\tiny $GBW$}}_{\!p}(x,k_{\!\perp}^2) \propto
k_{\!\perp}^2$ for small momenta. With increasing
$|\vec{k}_{\!\perp}|$, the non-perturbative contribution to ${\cal
  F}_{\!p}(x,k_{\!\perp}^2)$ decreases rapidly which results from the
strong suppression of large momenta by the non-perturbative
correlation functions $i\tilde{D}_1^{\prime \,(2)}(k_{\!\perp}^2)$ and
$i\tilde{D}^{(2)}(k_{\!\perp}^2)$ given
in~(\ref{Eq_D_(prime)(k^2)_for_exp_correlation})
and~(\ref{Eq_D(k^2)_for_exp_correlation}). For $|\vec{k}_{\!\perp}|
\,\gtsim\, 1\,\GeV$, the perturbative contribution dominates the
unintegrated gluon distribution. It drops as $1/k_{\!\perp}^2$ in
  accordance with the perturbative correlation function $i\tilde{D}_{\pert}^{\prime
  \,(2)}(k_{\!\perp}^2)$ given in~(\ref{Eq_massive_D_pge_prime_trans}).
This perturbative QCD result is not reproduced by the phenomenological
model of Golec-Biernat and W{\"u}sthoff~\cite{Golec-Biernat:1999qd}
which predicts a Gaussian decrease of ${\cal F}_{\!p}(x,k_{\!\perp}^2)$
with increasing $|\vec{k}_{\!\perp}|$.

The $x$\,-\,dependence of ${\cal F}_{\!p}(x,k_{\!\perp}^2)$ can be
seen in Figs.~\ref{fg_k.eps} and~\ref{p_fg_k_p_vs_np.eps}: With
decreasing $x$, the perturbative contribution increases much stronger
than the non-perturbative contribution which results from the energy
exponents $\epsilon^{\pert} \gg \epsilon^{\nprt}$ in~(\ref{x_dep})
necessary to describe the data within our model~\cite{Shoshi:2002in}.
Moreover, the perturbative contribution extends into the
small\,-\,$|\vec{k}_{\!\perp}|$ region as $x$ decreases. Indeed, the
soft-hard transition point moves towards smaller momenta with
decreasing $x$ as shown in Fig.~\ref{p_fg_k_p_vs_np.eps}. Such a
hard-to-soft diffusion is observed also in~\cite{Ivanov:2000cm} where
the unintegrated gluon distribution has been parametrized to reproduce
the experimental data for the proton structure function $F_2(x,Q^2)$
at small $x$. The opposite behavior is obtained in the approach of the
color glass condensate~\cite{Iancu:2002xk}: With decreasing $x$, gluons
are produced predominantely in the high\,-\,$|\vec{k}_{\!\perp}|$
region of lower density and weaker repulsive interactions.

\begin{figure}[htb]
\setlength{\unitlength}{1.cm}
\begin{center}
\epsfig{file=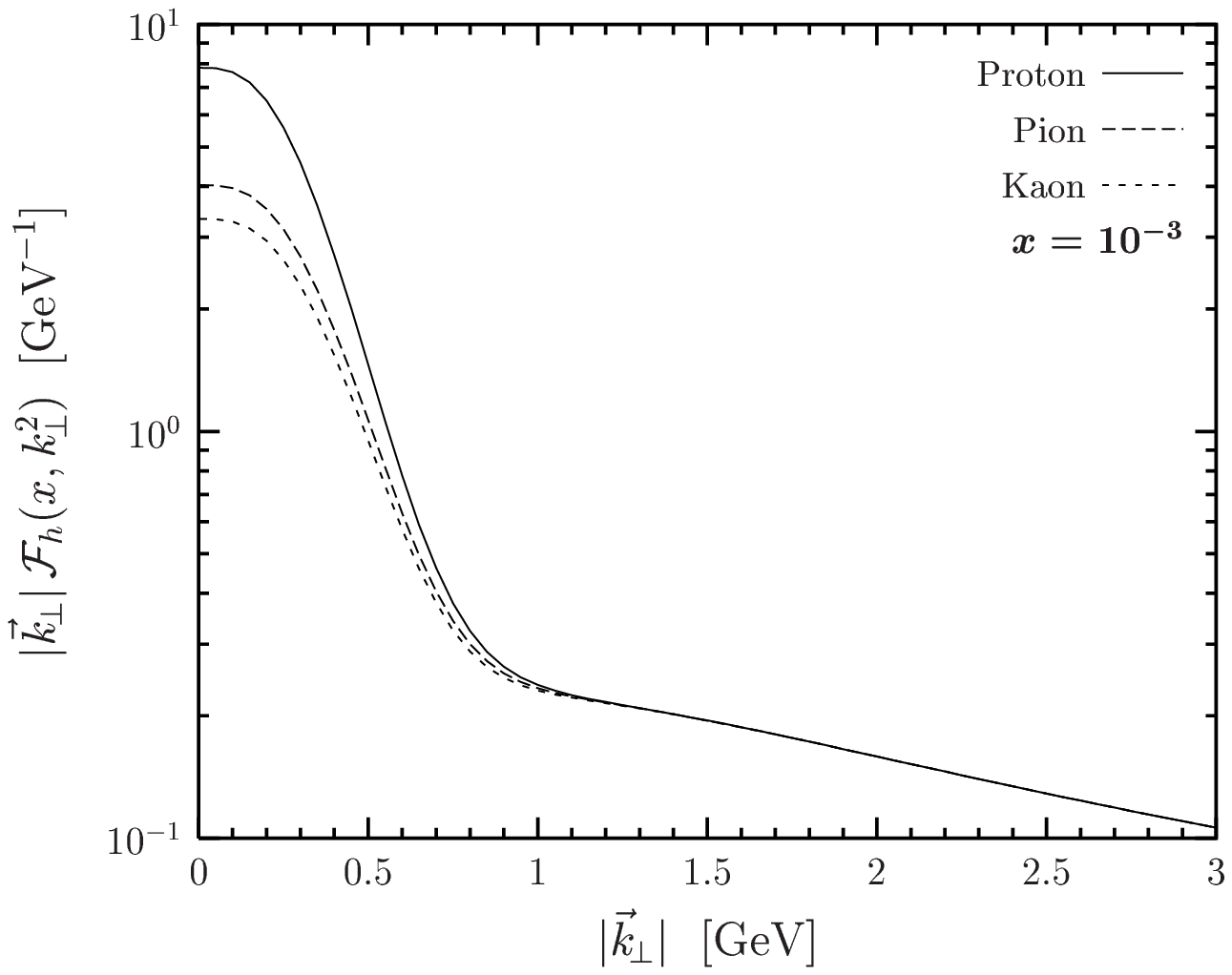,width=13.cm}
\end{center}
\caption{\small The unintegrated gluon distribution of the proton,
  pion, and kaon ${\cal F}_h(x,k_{\!\perp}^2)$ times the transverse
  momentum $|\vec{k}_{\!\perp}|$ as a function of
  $|\vec{k}_{\!\perp}|$ at Bjorken-variable $x=10^{-3}$.}
\label{p_pi_k_k_fg_k.eps}
\end{figure}
In Fig.~\ref{p_pi_k_k_fg_k.eps}, the unintegrated gluon distribution
of the proton, pion, and kaon ${\cal F}_h(x,k_{\!\perp}^2)$ times the
transverse momentum $|\vec{k}_{\!\perp}|$ is shown as a function of
$|\vec{k}_{\!\perp}|$ at $x=10^{-3}$. The hadrons are characterized by
different values for $\Delta z_h$ and $S_h$ in the hadron wave
function~(\ref{Eq_hadron_wave_function}). However, ${\cal
  F}_h(x,k_{\!\perp}^2)$ depends only on $S_h$. Due to the
normalization of the hadron wave functions, $\Delta z_h$ disappears
upon the integration over $z_i$ as can be seen directly
from~(\ref{averages_explicit_qq}) and~(\ref{averages_explicit_ss}). At
small momenta, ${\cal F}_h(x,k_{\!\perp}^2) \propto S_h^2$ is found
from~(\ref{averages_explicit_qq}), (\ref{averages_explicit_ss}),
and~(\ref{lf}). It becomes visible in Fig.~\ref{p_pi_k_k_fg_k.eps} for
the chosen hadron extensions: $S_p = 0.86\,\fm$, $S_{\pi} =
0.607\,\fm$, and $S_{K} = 0.55\,\fm$.  At large momenta where the
perturbative contribution dominates, the dependence on $S_h$ vanishes
as can be seen from~(\ref{averages_explicit_qq}) and the unintegrated
gluon distributions of protons, pions, and kaons become identical. Of
course, this behavior results from the wave function normalization
being identical for protons, pions, and kaons with two valence
constituents which are the quark and antiquark in the pion and kaon
and the quark and diquark in the proton. At large
$|\vec{k}_{\!\perp}|$, i.e., high resolution, the realistic
description of protons as three-quark systems becomes necessary. In
fact, the three-quark description of protons leads to a different
perturbative contribution in~(\ref{lf}): the quark counting factors of
$2$ (appropriate for mesons) in the square brackets in~(\ref{Ip}) are
substituted by factors of $3$ (appropriate for
baryons)~\cite{Low:1975sv+X,Gunion:iy}. At other values of $x$, the
unintegrated gluon distributions of protons, pions and kaons show the
same features.

\begin{figure}[htb]
\setlength{\unitlength}{1.cm}
\begin{center}
\epsfig{file=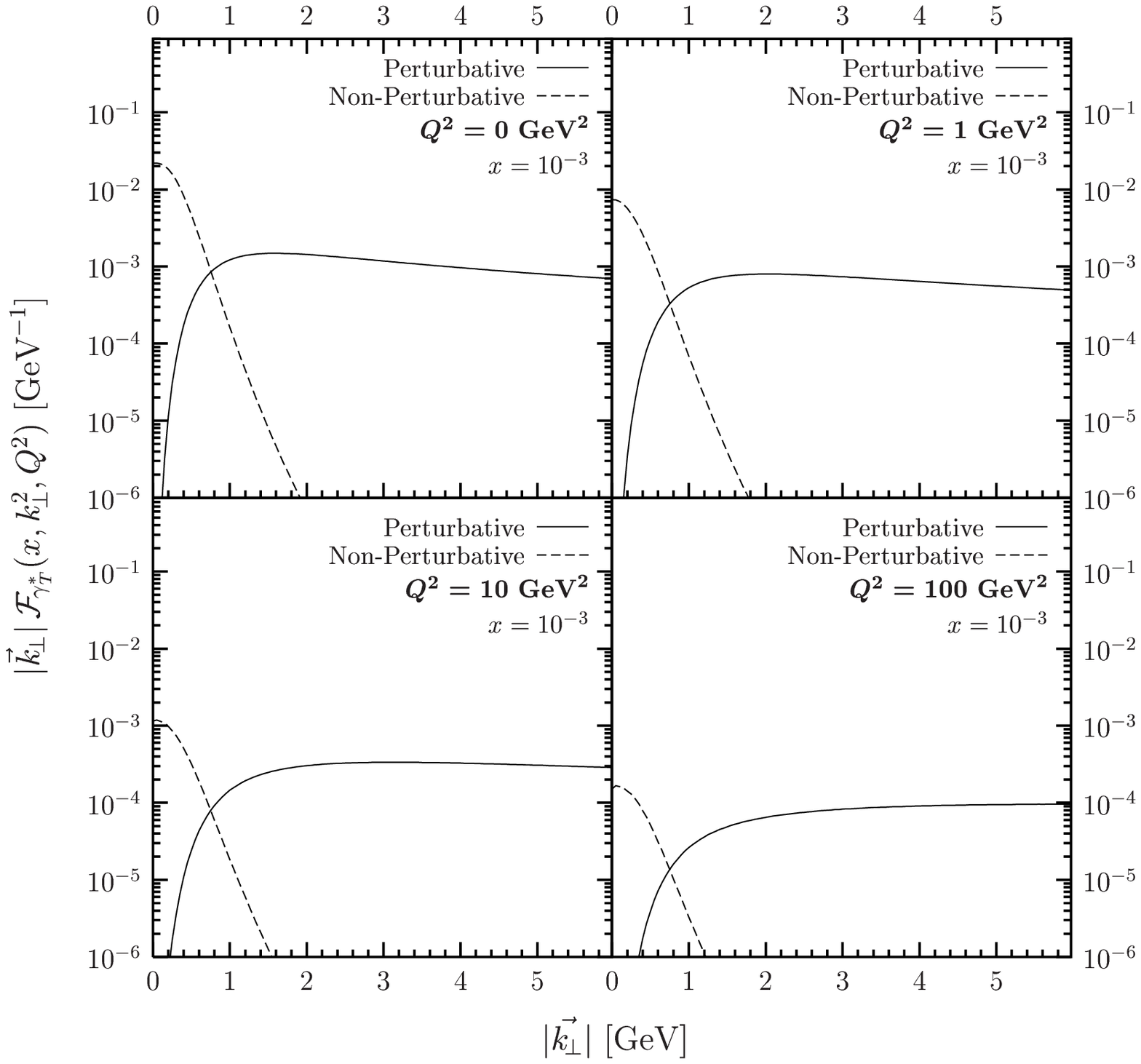,width=13.cm}
\end{center}
\caption{\small The unintegrated gluon distribution of the transverse
  polarized photon ${\cal F}_{\!\gamma^*_T}(x,k_{\!\perp}^2,Q^2)$
  times the tranverse momentum $|\vec{k}_{\!\perp}|$ as a function of
  $|\vec{k}_{\!\perp}|$ at photon virtualities of $Q^2 = 0$, $1$, $10$,
  and $100\,\GeV^2$ and Bjorken-variable $x=10^{-3}$.}
\label{gT_k_fg_k.eps}
\end{figure}
\begin{figure}[htb]
\setlength{\unitlength}{1.cm}
\begin{center}
\epsfig{file=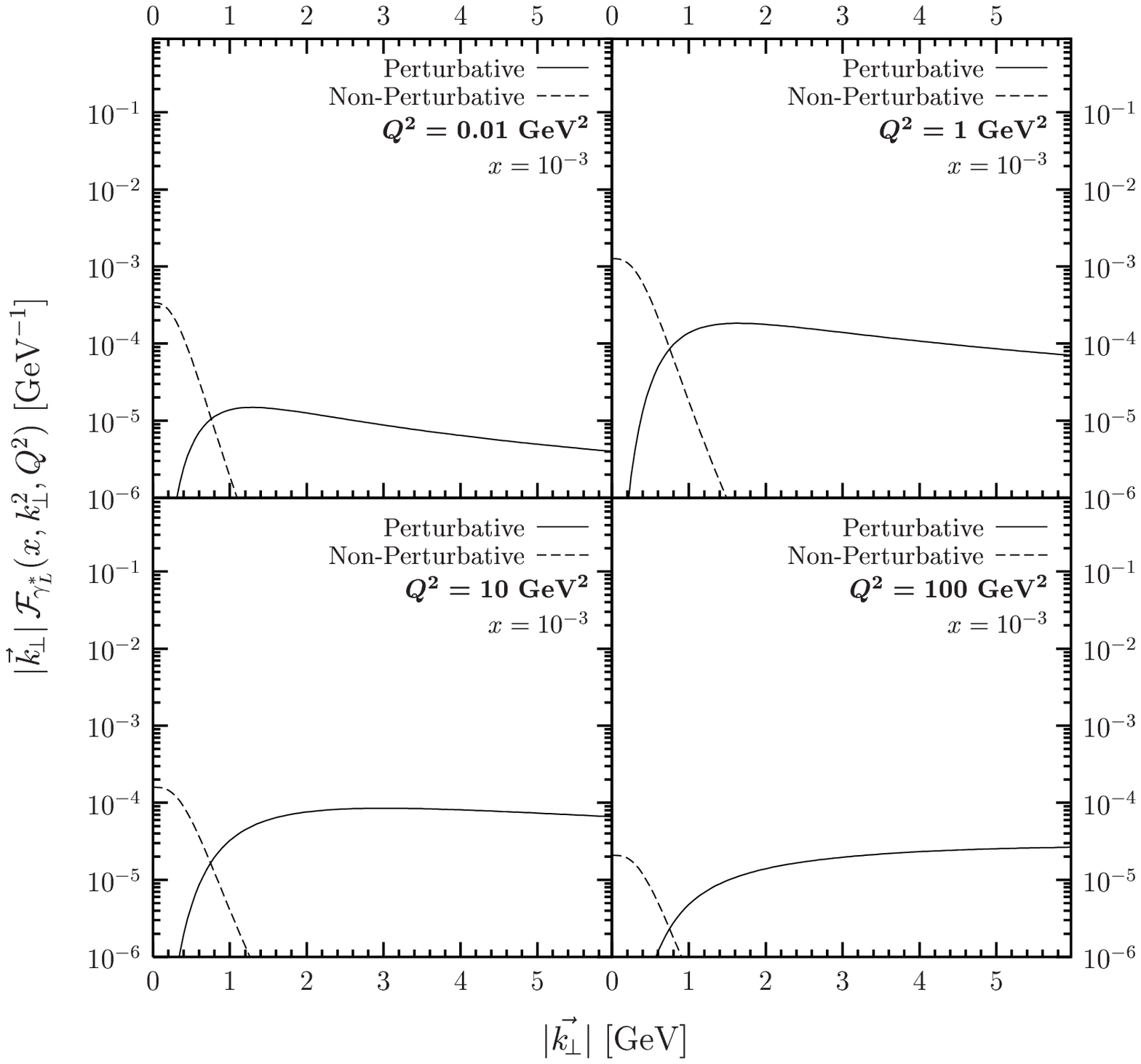,width=13.cm}
\end{center}
\caption{\small The unintegrated gluon distribution of the
  longitudinally polarized photon ${\cal
    F}_{\!\gamma^*_L}(x,k_{\!\perp}^2,Q^2)$ times the tranverse
  momentum $|\vec{k}_{\!\perp}|$ as a function of
  $|\vec{k}_{\!\perp}|$ at photon virtualities of $Q^2 = 0.01$, $1$,
  $10$, and $100\,\GeV^2$ and Bjorken-variable $x=10^{-3}$.}
\label{gL_k_fg_k.eps}
\end{figure}

Photons are particularly interesting because the
transverse size of the quark-antiquark dipole into which a photon
fluctuates is controlled by the photon virtuality~$Q^2$ (cf.
Appendix~\ref{Sec_Wave_Functions})
\be 
        |\vec{r}_{\gamma^*_{T,L}}| \approx
                         \frac{2}{Q^2+4m^2_u(Q^2)} \ , 
\label{rQ}
\ee 
where $m_u(Q^2)$ is the running $u$-quark mass given in
Appendix~\ref{Sec_Wave_Functions}. In Figs.~\ref{gT_k_fg_k.eps}
and~\ref{gL_k_fg_k.eps}, the unintegrated gluon distribution of
transverse~($T$) and longitudinal~($L$) photons ${\cal
  F}_{\!\gamma^*_{T,L}}(x,k_{\!\perp}^2,Q^2)$ times the transverse
momentum $|\vec{k}_{\!\perp}|$ is shown as a function of
$|\vec{k}_{\!\perp}|$ for various photon virtualities $Q^2$ at
$x=10^{-3}$.
With increasing $Q^2$, i.e., decreasing ``photon size''
$|\vec{r}_{\gamma^*_{T,L}}|$, the ratio of the perturbative to the
non-perturbative contribution to ${\cal
  F}_{\!\gamma^*_{T,L}}(x,k_{\!\perp}^2,Q^2)$ increases. This behavior
has already been discussed in Sec.~\ref{Momentum-Space Structure of
  Dipole-Dipole Scattering} on the level of dipole-dipole scattering,
where decreasing dipole sizes increase the ratio of the perturbative
to non-perturbative contribution, see
Fig.~\ref{Fig_dipole_dipole_interactions}. Due to the different
$Q^2$\,-\,dependence in the transverse and longitudinally polarized
photon wave functions, ${\cal
  F}_{\!\gamma^*_{T}}(x,k_{\!\perp}^2,Q^2)$ decreases continously with
increasing $Q^2$ while ${\cal
  F}_{\!\gamma^*_{L}}(x,k_{\!\perp}^2,Q^2)$ increases for $Q^2 \ltsim
1\,\GeV^2$ and decreases for $Q^2 \,\gtsim\, 1\,\GeV^2$.

\begin{figure}[htb]
\setlength{\unitlength}{1.cm}
\begin{center}
\epsfig{file=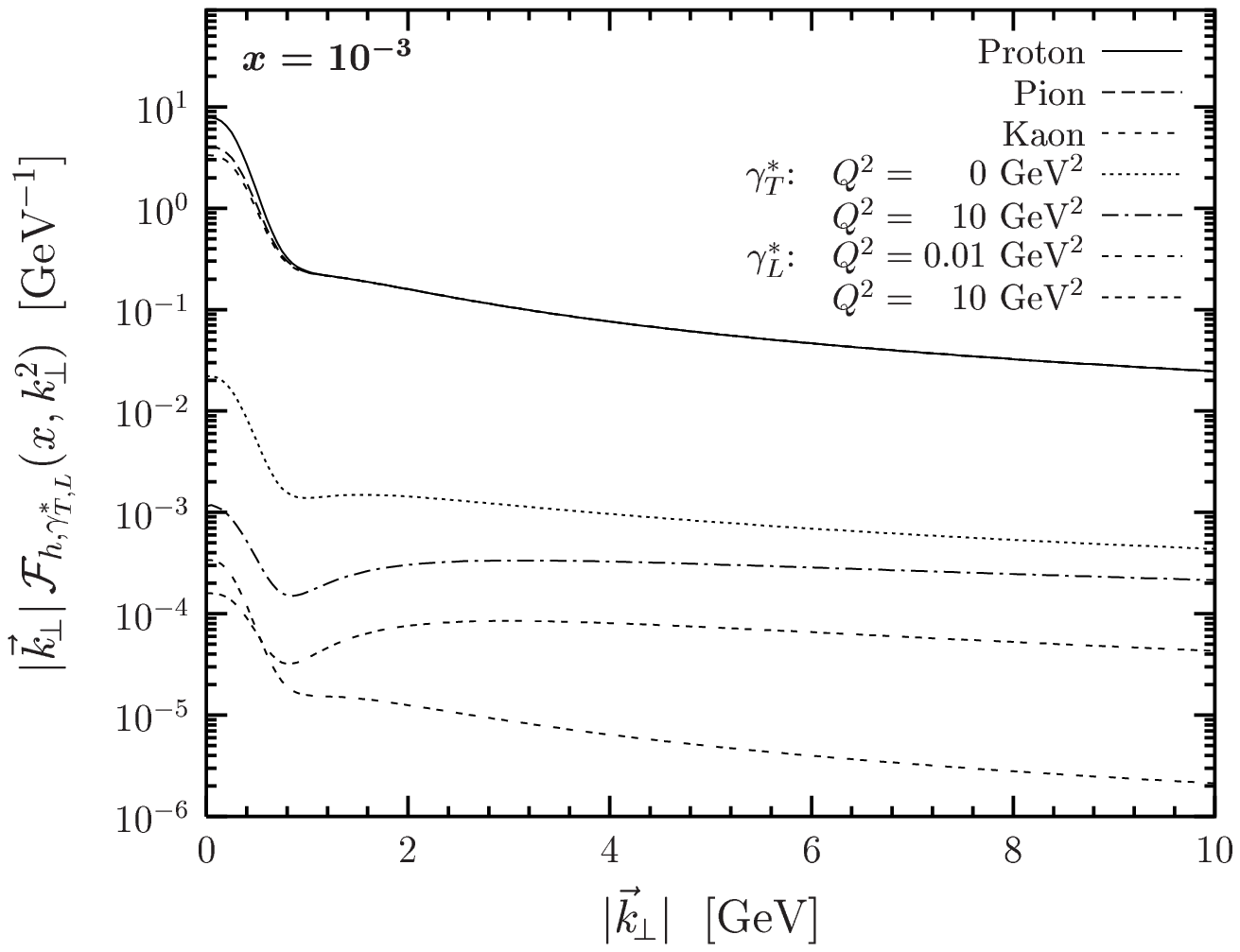,width=13.cm}
\end{center}
\caption{\small The unintegrated gluon distribution of the proton,
  pion, kaon, and transverse and longitudinally polarized photon
  ${\cal F}_{h,\gamma^*_{T,L}}(x,k_{\!\perp}^2)$ times the transverse
  momentum $|\vec{k}_{\!\perp}|$ as a function of
  $|\vec{k}_{\!\perp}|$ at Bjorken-variable $x=10^{-3}$. Results for
  transverse polarized photons are shown for photon virtualities of
  $Q^2=0$ and $10\,\GeV^2$ and results for longitudinally polarized
  photons for photon virtualities of $Q^2=0.01$ and $10\,\GeV^2$.}
\label{h_g_k_fg_k.eps}
\end{figure}
In Fig.~\ref{h_g_k_fg_k.eps}, the unintegrated gluon distributions of
hadrons and photons discussed above are compared. The unintegrated
gluon distribution of real ($Q^2=0$) photons ($|\vec{r}_{\gamma^{}_T}|
\approx S_h$) is suppressed by a factor of order $\alpha$ in
comparison to the one of the hadrons otherwise its shape is very
similar. The suppression factor comes from the photon\,-\,dipole
transition described by the photon wave functions given in
Appendix~\ref{Sec_Wave_Functions}. The ratio of the perturbative to
the non-perturbative contribution of unintegrated gluon distributions
increases as one goes from hadrons to photons with high virtuality
$Q^2$. Since the wave functions of protons, pions, and kaons are
normalized to the same value, the unintegrated gluon distributions of
these hadrons are, as mentioned above, identical at large
$|\vec{k}_{\!\perp}|$ and do not depend on the hadron size. In
contrast, the $Q^2$\,-\,dependence of the photon wave functions leads
to $Q^2$\,-\,dependent, i.e., ``photon size''\,-\,dependent,
unintegrated gluon distributions at large $|\vec{k}_{\!\perp}|$. With
increasing $|\vec{k}_{\!\perp}|$, the unintegrated gluon distributions
of hadrons and photons become parallel in line with the vanishing
dependence on the specific form of the wave function.

\begin{figure}[htb]
\setlength{\unitlength}{1.cm}
\begin{center}
\epsfig{file=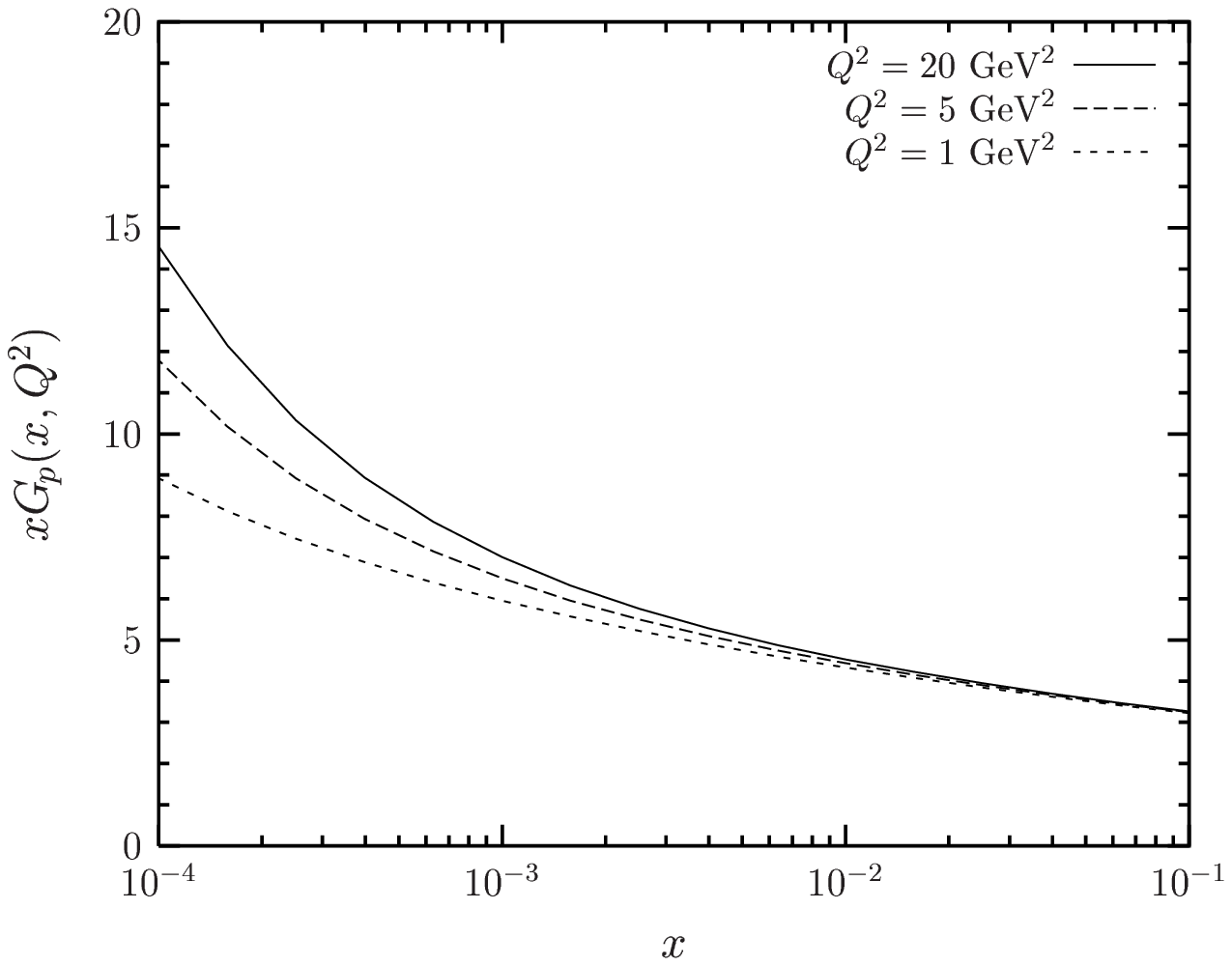,width=13.cm}
\end{center}
\caption{\small The gluon distribution of the proton
  $xG_p(x,Q^2)$ as a function of Bjorken\,-\,$x$ at photon
  virtualities of $Q^2=1$, $5$, and $20\, \GeV^2$.}
\label{xG_x.eps}
\end{figure}
In Fig.~\ref{xG_x.eps}, the integrated gluon distribution of the
proton $xG_p(x,Q^2)$ is shown as a function of Bjorken\,-\,$x$ at
photon virtualities of $Q^2 = 1, 5$, and $20\,\GeV^2$.  Recall that
the parameter $x_0=2.4 \cdot 10^{-3}$ has been adjusted in the
previous section such that the experimental data of $xG_p(x,Q^2)$ at
$Q^2=1\,\GeV^2$~\cite{Abramowicz:1998ii+X} are reproduced. For $x
\,\gtsim\, 10^{-3}$, $xG_p(x,Q^2)$ is mainly determined by
non-perturbative physics as can be seen from Figs.~\ref{fg_k.eps}
and~\ref{p_fg_k_p_vs_np.eps}. Perturbative physics becomes relevant
for $x\,\ltsim\,10^{-3}$ and generates the steep increase of
$xG_p(x,Q^2)$ with decreasing $x$ at fixed $Q^2$. Also the rise of
$xG_p(x,Q^2)$ with increasing $Q^2$ at fixed $x$ results from the
perturbative contribution. For $x \ll 10^{-4}$, we have shown
explicitly in our recent work~\cite{Shoshi:2002in,Shoshi:2002ri} that
multiple gluonic exchanges contained in the full $T$-matrix
element~(\ref{Eq_model_purely_imaginary_T_amplitude_almost_final_result})
slow down the powerlike increase of $xG_p(x,Q^2)$ with decreasing $x$
in accordance with $S$-matrix unitarity constraints.

\newpage
\section{Comparison with Other Work}
\label{Sec_Comparison_with_Other_Work}

In this section, we compare the unintegrated gluon distribution of the
proton extracted from our loop-loop correlation model (LLCM) with
those obtained from the saturation model of Golec-Biernat and
W{\"u}sthoff (GBW)~\cite{Golec-Biernat:1999qd}, the derivative of the
Gl{\"u}ck, Reya, and Vogt (GRV) parametrization of
$xG_p(x,Q^2)$~\cite{Gluck:1998xa}, and the approach of Ivanov and
Nikolaev (IN)~\cite{Ivanov:2000cm}.

\medskip

In the approach of Golec-Biernat and
W{\"u}sthoff~\cite{Golec-Biernat:1999qd}, the unintegrated gluon
distribution of the proton is extracted from the total dipole-proton
cross section by inverting~(\ref{unint_gl_distr_def}),
\be
{\cal F}_{p}(x,k_{\!\perp}^2) = 
        \frac{3\,\sigma_0}{4\,\pi^2\,\alphaS}\,
        R_0^2(x)\,k^2_{\!\perp}\,\exp\left(-R_0^2(x)\,k^2_{\!\perp}\right) 
\quad \mbox{with} \quad
        R_0^2(x) = \frac{1}{Q_0^2}\left(\frac{x}{x_0}\right)^{\lambda}
\label{GBW}
\ee
where the parameter $\sigma_0 = 29.12\,\mb$, $\alphaS = 0.2$, $Q_0 =
1\,\GeV$, $\lambda = 0.277$, and $x_0 = 0.41 \cdot 10^{-4}$ are
obtained from a fit to the proton structure function $F_2(x,Q^2)$
including charm quarks in the photon wave function. Note, however that
the GBW approach uses the dipole-proton cross section of the GBW model
on the lhs of~(\ref{unint_gl_distr_def}) which implies multiple gluon
exchanges while the rhs of~(\ref{unint_gl_distr_def}) describes only
two-gluon exchange as discussed in Sec.~\ref{Momentum-Space Structure
  of Dipole-Dipole Scattering}. Moreover, as demonstrated below, the
large $k^2_{\!\perp}$\,-\,behavior of the unintegrated gluon
distribution~(\ref{GBW}) deviates significantly from the DGLAP
results. This mismatch motivated the recent modifications of the
saturation model~\cite{Bartels:2002cj}.

\medskip

The unintegrated gluon distribution of the proton is also computed
from the integrated gluon distribution $xG_p(x,Q^2)$ by
inverting~(\ref{Eq_xG(x,Q^2)})
\be
{\cal F}_{p}(x,k_{\!\perp}^2) =
        \left.\frac{dxG_p(x,Q^2)}{dQ^2}\right|_{Q^2=k^2_{\!\perp}} \ .
\label{GRV}
\ee
Here, we use for the integrated gluon distribution $xG_p(x,Q^2)$ the
leading order (LO) parametrization of Gl{\"u}ck, Reya, and Vogt
(GRV)~\cite{Gluck:1998xa}, which covers the kinematic region $10^{-9}
< x < 1$ and $0.8\,\GeV^2 < Q^2 < 10^6\,\GeV^2$.

\medskip

Ivanov and Nikolaev~\cite{Ivanov:2000cm} have constructed a
two-component (soft $+$ hard) ansatz for the unintegrated gluon
distribution of the proton
\be
{\cal F}_p(x,k_{\!\perp}^2) 
        = {\cal
          F}_{\soft}(x,k_{\!\perp}^2)\,\frac{k^2_s}{k^2_{\!\perp}+k^2_s} 
        + {\cal
         F}_{\hard}(x,k_{\!\perp}^2)\,\frac{k^2_{\!\perp}}{k^2_{\!\perp} +
         k^2_h}
\label{IN}
\ee
with the soft and hard component
\bea
\!\!\!\!\!\!\!\!
{\cal F}_{\soft}(k_{\!\perp}^2) 
        & = &
       a_{\soft}\,C_F\,N_c\,\frac{\alphaS(k^2_{\!\perp})}{\pi}\,
       \frac{k^2_{\!\perp}}{\left(k^2_{\!\perp}+\mu_{\soft}^2\right)^2}\,
       \left [1-
         \left(1+\frac{3\,k^2_{\!\perp}}{\Lambda^2}\right)^{-2}\right ]
\label{F_IN_soft}\\
\!\!\!\!\!\!\!\!
{\cal F}_{\hard}(x,k_{\!\perp}^2) 
        & = &
      {\cal F}^{(B)}_{\pt}(k_{\!\perp}^2)\,\frac{{\cal
      F}_{\pt}(x,Q^2_c)}{{\cal
      F}_{\pt}^{(B)}(Q^2_c)}\,\Theta(Q^2_c-k^2_{\!\perp}) +
      {\cal F}_{\pt}(x,k^2_{\!\perp})\,\Theta(k^2_{\!\perp}-Q^2_c) 
\label{F_IN_hard}
\eea
where ${\cal F}^{(B)}_{\pt}(k_{\!\perp}^2)$ has the same form
as~(\ref{F_IN_soft}) with the parameters $a_{\pt}$ and $\mu_{\pt}$
instead of $a_{\soft}$ and $\mu_{\soft}$ and ${\cal
  F}_{\pt}(x,k^2_{\!\perp})$ is the derivative of the integrated gluon
distribution~(\ref{GRV}). In the IN approach, the running coupling is
given by $\alphaS(k^2_{\!\perp}) = \mbox{min}\{0.82,
(4\,\pi)/(\beta_0\,\log[k^2_{\!\perp}/\Lambda^2_{QCD}])\}$. With the
GRV-parametrization~\cite{Gluck:1998xa} for the integrated gluon
distribution, the structure function of the proton $F_2(x,Q^2)$ has
been described successfully using the following parameters: $k^2_s =
3\,\GeV^2$, $k^2_h = (1+0.0018\log(1/x)^4)^{0.5}$, $a_{\soft}=2$,
$a_{\pt}=1$, $\mu_{\soft}=0.1\,\GeV$, $\mu_{\pt}=0.75\,\GeV$, $Q^2_c =
0.895\,\GeV^2$, $\beta_0 = 9$, and $\Lambda_{QCD} = 0.2\,\GeV$.

\medskip

\begin{figure}[p!]
  \setlength{\unitlength}{1.cm}
\begin{center}
\epsfig{file=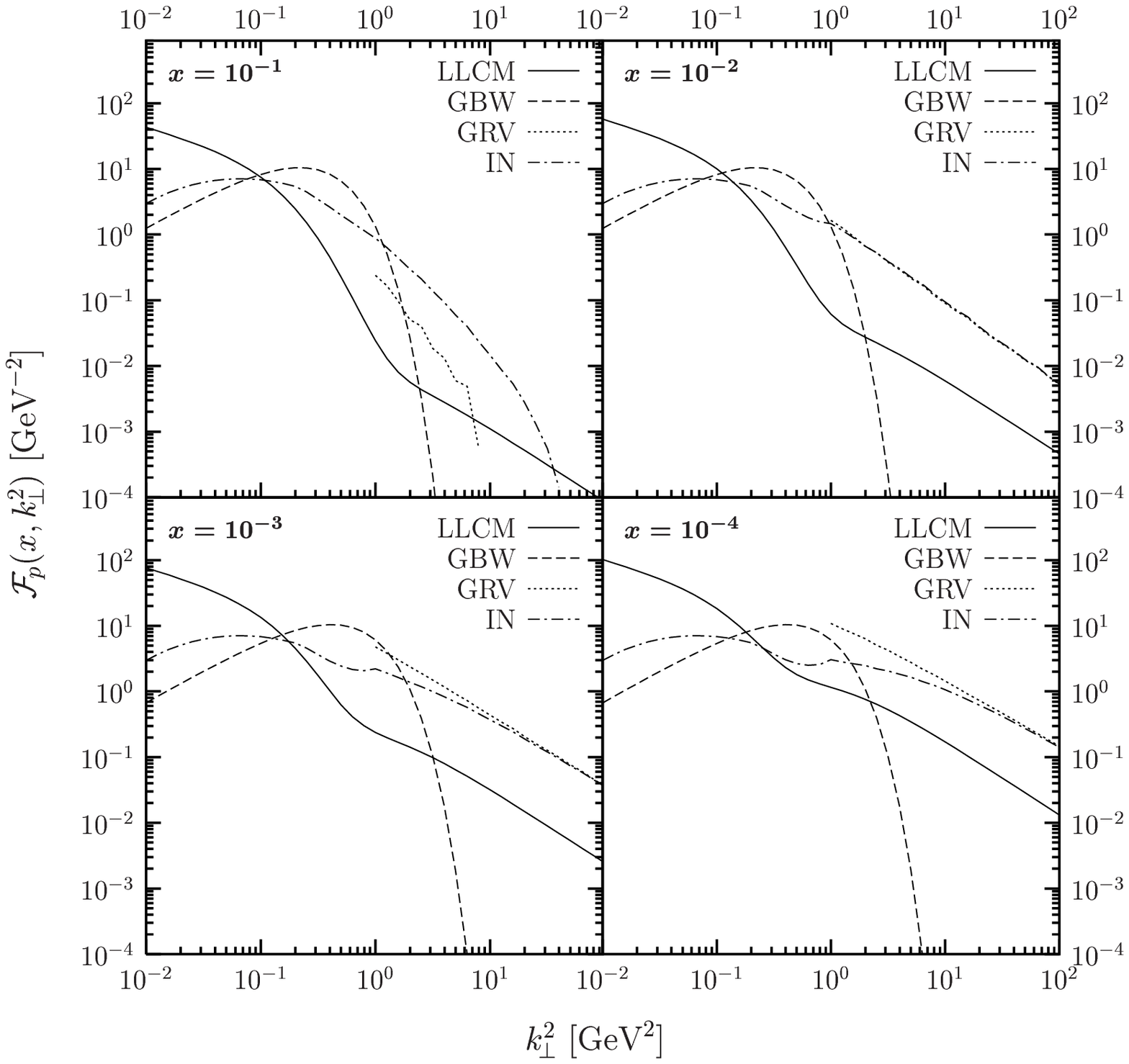,width=15.cm}
\end{center}
\caption{\small The unintegrated gluon distribution of the
  proton ${\cal F}_{p}(x,k_{\!\perp}^2)$ as a function of transverse
  momentum squared $k^2_{\!\perp}$ at Bjorken\,-\,$x$ values of
  $10^{-1}$, $10^{-2}$, $10^{-3}$, and $10^{-4}$. The different curves
  are obtained from our loop-loop correlation model (LLCM), the
  Golec-Biernat and W{\"u}sthoff (GBW)
  model~\cite{Golec-Biernat:1999qd}, the derivative of the Gl{\"u}ck,
  Reya, and Vogt (GRV) parametrization of
  $xG_p(x,Q^2)$~\cite{Gluck:1998xa}, and the Ivanov and Nikolaev (IN)
  approach~\cite{Ivanov:2000cm}.}
\label{fg_SSDP_GBW_IN_GRV_k2_4plots.eps}
\end{figure}
In Fig.~\ref{fg_SSDP_GBW_IN_GRV_k2_4plots.eps}, we show the LLCM, GBW,
GRV, and IN results for the unintegrated gluon distribution of the
proton ${\cal F}_{p}(x,k_{\!\perp}^2)$ as a function of transverse
momentum squared $k^2_{\!\perp}$ for $x=10^{-1}$, $10^{-2}$,
$10^{-3}$, and $10^{-4}$. At small transverse momenta,
$k_{\!\perp}^2\,\ltsim\,0.1\,\GeV^2$, our model gives the largest
values for ${\cal F}_{p}(x,k_{\!\perp}^2)$. As mentioned in the
previous section, our LLCM unintegrated gluon distribution increases
as $1/\sqrt{k_{\!\perp}^2}$ with decreasing $k_{\!\perp}^2$ as a
consequence of the linear increase of the total dipole-proton cross
section at large dipole sizes. In contrast, for $k_{\!\perp}^2 \to 0$,
the unintegrated gluon distribution of GBW decreases as
$k_{\!\perp}^2$ and the one of IN as $k_{\!\perp}^4$. In the
perturbative region, $k_{\!\perp}^2\,\gtsim\,1\,\GeV^2$, the
unintegrated gluon distribution of the LLCM becomes smaller than the
one of GRV and IN but is still larger than the one of GBW. Moreover,
the LLCM, GRV, and IN unintegrated gluon distributions become parallel
for $x\,\ltsim\,10^{-2}$ and drop as $1/k_{\!\perp}^2$ for large
$k_{\!\perp}^2$. This perturbative QCD behavior is not reproduced by
the GBW unintegrated gluon distribution which decreases exponentially
with increasing $k_{\!\perp}^2$.

\medskip

\begin{figure}[p!]
\setlength{\unitlength}{1.cm}
\begin{center}
\epsfig{file=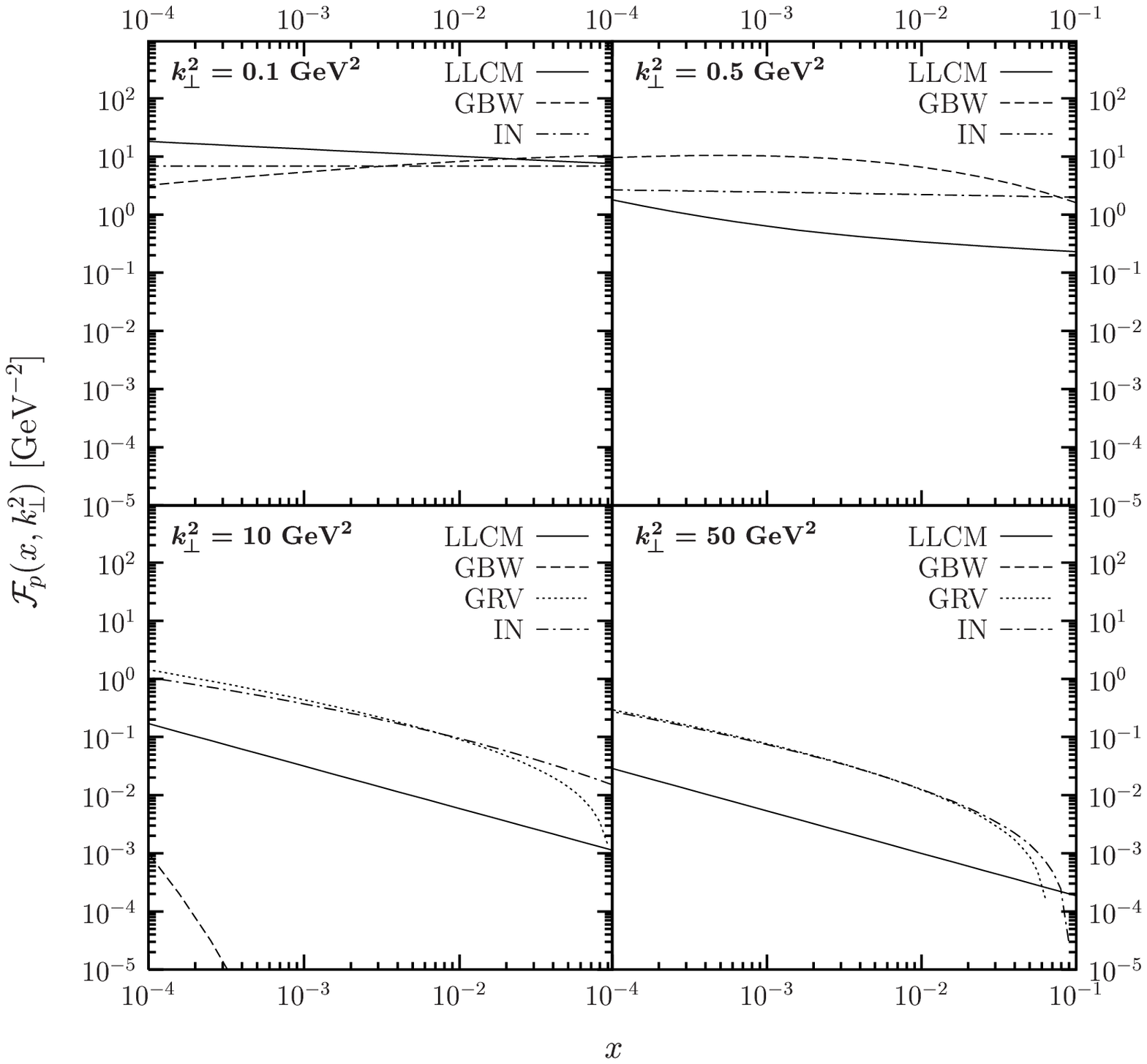,width=15.cm}
\end{center}
\caption{\small The unintegrated gluon distribution of the
  proton ${\cal F}_{p}(x,k_{\!\perp}^2)$ as a function of
  Bjorken\,-\,$x$ at transverse momenta squared $k^2_{\!\perp}=0.1$,
  $0.5$, $10$, and $50\,\GeV^2$. The different curves are obtained
  from our loop-loop correlation model (LLCM), the Golec-Biernat and
  W{\"u}sthoff (GBW) model~\cite{Golec-Biernat:1999qd}, the derivative
  of the Gl{\"u}ck, Reya, and Vogt (GRV) parametrization of
  $xG_p(x,Q^2)$~\cite{Gluck:1998xa}, and the Ivanov and Nikolaev (IN)
  approach~\cite{Ivanov:2000cm}. Note that the GRV parametrization is
  only available for $k^2_{\!\perp} \ge 0.8\,\GeV^2$. Moreover, the
  GBW result for ${\cal F}_{p}(x,k_{\!\perp}^2)$ is below $10^{-5}$
  for $k^2_{\!\perp} = 50\,\GeV^2$.}
\label{fg_SSDP_GBW_IN_GRV_x_4plots.eps}
\end{figure}
The $x$-dependence of the LLCM, GBW, GRV, and IN unintegrated gluon
distributions ${\cal F}_{p}(x,k_{\!\perp}^2)$ is shown for transverse
momenta squared $k^2_{\!\perp}=0.1$, $0.5$, $10$, and $50\,\GeV^2$ in
Fig.~\ref{fg_SSDP_GBW_IN_GRV_x_4plots.eps}.  In the non-perturbative
region, $k_{\!\perp}^2\,=\,0.1\,\GeV^2$ and
$k_{\!\perp}^2\,=\,0.5\,\GeV^2$, the LLCM, GBW, and IN unintegrated
gluon distributions show a weak $x$-dependence.  In the perturbative
region, $k_{\!\perp}^2\,=\,10\,\GeV^2$ and
$k_{\!\perp}^2\,=\,50\,\GeV^2$, the $x$-dependence of the unintegrated
gluon distributions becomes stronger. The LLCM, GRV, and IN
unintegrated gluon distributions show nearly the same rise with
decreasing $x$. In contrast, the GBW unintegrated gluon distribution
increases much faster as $x$ decreases.

In addition to the one-scale unintegrated gluon distributions, ${\cal
  F}(x,k_{\!\perp}^2)$, discussed in this work, there exist also
two-scale unintegrated gluon distributions, ${\cal
  F}(x,k_{\!\perp}^2,\mu^2)$. In the CCFM evolution
equation~\cite{Ciafaloni:1987ur+X}, this additional scale $\mu^2$ is
related to the maximum angle allowed in the gluon emission. Two-scale
unintegrated gluon distributions are obtained in the approach of
Bl\"umlein~\cite{Blumlein:1995eu}, Jung and
Salam~\cite{Jung:1998mi+X}, Kimber, Martin, and
Ryskin~\cite{Kimber:2001sc}, and in the linked dipole chain (LDC)
model~\cite{Andersson:1995ju+X,Gustafson:2002jy}. A comparison of
their results can be found in~\cite{Anderson:2002cf,Gustafson:2002jy}
where also the one-scale unintegrated gluon distributions of
Kwiecinski, Martin and Stasto~\cite{Kwiecinski:1997ee}, and Ryskin and
Shabelski~\cite{Ryskin:wz} are discussed.

\section{Conclusions}
\label{Sec_Conclusions}

We have investigated perturbative and non-perturbative QCD
contributions to high-energy scattering in momentum space within the
recently presented loop-loop correlation model~\cite{Shoshi:2002in}.
The perturbative contribution has been described by two-gluon exchange
and the non-perturbative contribution by the stochastic vacuum model
of QCD which leads to confinement of the quark and antiquark in a
dipole via a string of color fields. This QCD string gives important
non-perturbative contributions to high-energy reactions and manifests
itself in the linear increase of the total dipole-hadron cross section
for large dipole sizes. A new structure different from perturbative
two-gluon exchange has been found in the non-perturbative
string-string interaction. A decomposition of the confining string
into dipoles has allowed us to extract the microscopic structure of
the unintegrated gluon distribution of hadrons and photons from the
dipole-hadron and dipole-photon cross section via
$|\vec{k}_{\!\perp}|$\,-\,factorization.

The minimal surfaces used in our loop-loop correlation
model~\cite{Shoshi:2002in} have allowed us to show for the first time
the QCD structure of the non-perturbative contribution to the
dipole-dipole scattering amplitude in momentum space. This
contribution has two parts: The first part describes the
non-perturbative interaction between the quarks and antiquarks of the
two dipoles and has the structure known from perturbative
two-gluon exchange~\cite{Low:1975sv+X,Gunion:iy}. The second part
describes the interaction between the strings of the two dipoles and
has a new structure originating from the geometry of the strings.
The string contribution, however, is negligible in the scattering of
two small dipoles which is governed by perturbative physics.

The contribution of the confining string to the total dipole-hadron
cross section $\sigma_{\!Dh}(x, |\vec{r}_{\!\mbox{\tiny\it D}}|)$ has
been studied. For small dipole sizes, $|\vec{r}_{\!\mbox{\tiny\it D}}|
\to 0$, the string contribution shows color transparency,
$\sigma_{\!Dh}(x,|\vec{r}_{\!\mbox{\tiny\it D}}|) \propto
r_{\!\mbox{\tiny\it D}}^2$, as known for the perturbative
contribution.  For large dipole sizes, $|\vec{r}_{\!\mbox{\tiny\it
    D}}|\,\gtsim\, 0.5\,\fm$, the non-perturbative contribution
increases linearly with increasing dipole size,
$\sigma_{\!Dh}(x,|\vec{r}_{\!\mbox{\tiny\it D}}|) \propto
|\vec{r}_{\!\mbox{\tiny\it D}}|$, in contrast to the perturbative
contribution which gives an $|\vec{r}_{\!\mbox{\tiny\it
    D}}|$\,-\,independent dipole-hadron cross section. This linear
increase is generated by the interaction of the string of the dipole
with the hadron: the longer the string, the larger the geometric cross
section with the hadron. String breaking, expected to stop the linear
increase for $|\vec{r}_{\!\mbox{\tiny\it D}}| \,\gtsim\, 1\, \fm$, is
not obtained in our model working in the quenched approximation. In
our model, $|\vec{k}_{\!\perp}|$\,-\,factorization has been found to be valid
also for the non-perturbative contribution to
$\sigma_{\!Dh}(x,|\vec{r}_{\!\mbox{\tiny\it D}}|)$.

A very interesting feature of the string confining the quark and
antiquark in the dipole has been found: A string of length
$|\vec{r}_{\!\mbox{\tiny\it D}}|$ can be represented as an integral
over stringless dipoles of sizes $\xi|\vec{r}_{\!\mbox{\tiny\it D}}|$
with $0 \leq \xi \leq 1$ and dipole number density $n(\xi) = 1/\xi^2$.
A similar behavior has been observed for the perturbative wave
function of a $q{\bar q}$ onium state in the large\,-\,$N_c$ limit
where the numerous emitted gluons are considered as
dipoles~\cite{Mueller:1994rr,Mueller:1994jq}. The decomposition of the
string into stringless dipoles has allowed us to rewrite the
string-hadron scattering process as an incoherent superposition of
dipole-hadron scattering processes and to extract the unintegrated
gluon distribution of hadrons and photons from our dipole-hadron and
dipole-photon cross section via
$|\vec{k}_{\!\perp}|$\,-\,factorization.

We have shown explicitly the microscopic structure of the unintegrated
gluon distribution in hadrons and photons. For small momenta
$|\vec{k}_{\!\perp}|$, the unintegrated gluon distributions of
protons, pions, and kaons ${\cal F}_h(x,k_{\!\perp}^2)$ are dominated
by non-perturbative physics and behave as $S_h^2/|\vec{k}_{\!\perp}|$
where $S_h$ denotes the hadron extension.  The
$1/|\vec{k}_{\!\perp}|$\,-\,behavior reflects the linear increase of
the total dipole-hadron cross section at large dipole sizes. The
unintegrated gluon distributions of protons, pions, and kaons differ
at small momenta because of their different extensions: $S_p =
0.86\,\fm$, $S_{\pi} = 0.607\,\fm$, and $S_{K} = 0.55\,\fm$.  For
large momenta, $|\vec{k}_{\!\perp}| \,\gtsim\, 1\,\GeV$, the
unintegrated gluon distributions of the hadrons are dominated by
perturbative physics and show the $1/k_{\!\perp}^2$\,-\,behavior
induced by the gluon propagator. In the perturbative region of large
momenta, the valence constituents are resolved and the dependence of
the unintegrated gluon distribution on the hadron extension $S_h$
vanishes. In contrast, the unintegrated gluon distribution of photons
depends on the ``photon size'' controlled by the photon virtuality
$Q^2$ for large $|\vec{k}_{\!\perp}|$. For very large
$|\vec{k}_{\!\perp}|$, the unintegrated gluon distributions of hadrons
and photons become parallel in line with the vanishing dependence on
the specific form of the wave function.

The $x$\,-\,dependence of the unintegrated gluon distribution of
hadrons and photons has been introduced phenomenologically into our
model. Motivated by the successful description of many experimental
data in our recent work~\cite{Shoshi:2002in}, we have given a strong
energy dependence to the perturbative contribution and a weak one to
the non-perturbative contribution. Consequently, with decreasing $x$
the perturbative contribution increases much stronger than the
non-perturbative contribution and extends into the
small\,-\,$|\vec{k}_{\!\perp}|$ region. A similar hard-to-soft
diffusion is observed also in the approach of Ivanov and
Nikolaev~\cite{Ivanov:2000cm} while a soft-to-hard diffusion is
obtained in the approach of the color glass
condensate~\cite{Iancu:2002xk}. Considering the integrated gluon
distribution of the proton $xG_p(x,Q^2)$, our non-perturbative
contribution dominates for $x \,\gtsim\, 10^{-3}$ while the
perturbative contribution becomes relevant for $x \,\ltsim\, 10^{-3}$
and generates the steep increase of $xG_p(x,Q^2)$ with decreasing $x$
at fixed $Q^2$. Also the rise of $xG_p(x,Q^2)$ with increasing $Q^2$
at fixed $x$ results from the strong energy dependence of the
perturbative contribution.

We have compared the unintegrated gluon distribution of the proton
${\cal F}_{p}(x,k_{\!\perp}^2)$ extracted from our loop-loop
correlation model (LLCM) with those obtained from the saturation model
of Golec-Biernat and W{\"u}sthoff (GBW)~\cite{Golec-Biernat:1999qd},
the derivative of the Gl{\"u}ck, Reya, and Vogt (GRV) parametrization
of $xG_p(x,Q^2)$~\cite{Gluck:1998xa}, and the approach of Ivanov and
Nikolaev (IN)~\cite{Ivanov:2000cm}. For $k_{\!\perp}^2 \to 0$, the
unintegrated gluon distribution of GBW decreases as $k_{\!\perp}^2$
and the one of IN as $k_{\!\perp}^4$ in contrast to the
$1/\sqrt{k_{\!\perp}^2}$\,-\,decrease found in our model.  In the
perturbative region, the LLCM, GRV, and IN unintegrated gluon
distributions become parallel for $x\,\ltsim\,10^{-2}$ and drop as
$1/k_{\!\perp}^2$ with increasing $k_{\!\perp}^2$. This perturbative
QCD behavior is not reproduced by the GBW unintegrated gluon
distribution which decreases exponentially with increasing
$k_{\!\perp}^2$. The $x$-dependence of the considered unintegrated
gluon distributions is weak in the non-perturbative region and becomes
stronger as $k_{\!\perp}^2$ increases.

The unintegrated gluon distributions of hadrons and photons, ${\cal
  F}_{h}(x,k_{\!\perp}^2)$, extracted from our model can be used to
compute, for example, dijet and vector meson production. So far, the
unintegrated gluon distribution is obtained from the lowest order
contribution to the $T$-matrix element and can only be used for
$x\,\gtsim\,10^{-4}$. For smaller values of $x$, multiple gluonic
exchanges become important and necessary to respect $S$-matrix
unitarity constraints~\cite{Shoshi:2002in}. Thus, a formalism must be
developed that allows to extract an unintegrated gluon distribution
which takes into account multiple gluonic exchanges.


\section*{Acknowledgements}

A.~Shoshi and F.~Steffen would like to thank C.~Ewerz, H.~Forkel, and
A.~Polleri for interesting discussions and Yu.~Ivanov for his support
in computational issues.  This research was supported in part by the
National Science Foundation under Grant No.~PHY99-07449 and the Intas
project ``Nonperturbative QCD.''

%
\begin{appendix}

\section{Hadron and Photon Wave Functions}
\label{Sec_Wave_Functions}

The light-cone wave functions $\psi_i(z_i,\vec{r}_i)$ provide the
distribution of transverse size and orientation ${\vec r}_{i}$ and
longitudinal quark momentum fraction $z_i$ to the light-like
Wegner-Wilson loops $W[C_i]$ that represent the scattering
color-dipoles. 

\subsection*{The Hadron Wave Function}

In this work, mesons and baryons are assumed to have a quark-antiquark
and quark-diquark valence structure, respectively.  This allows us to
model hadrons as color-dipoles.  We use for the hadron wave function
the phenomenological Gaussian Wirbel-Stech-Bauer
ansatz~\cite{Wirbel:1985ji}
\be
        \psi_h(z_i,\vec{r}_i) 
        = \sqrt{\frac{z_i(1-z_i)}{2 \pi S_h^2 N_h}}\, 
        e^{-(z_i-\inv{2})^2 / (4 \Delta z_h^2)}\,  
        e^{-|\vec{r}_i|^2 / (4 S_h^2)} 
        \ ,
\label{Eq_hadron_wave_function}
\ee
where the constant $N_h$ is fixed by the normalization
of~(\ref{Eq_hadron_wave_function}) to unity. The different hadrons
considered --- protons, pions, and kaons --- are characterized by
different values for $\Delta z_h$ and $S_h$. The extension parameter
$S_h$ is a fit parameter that should resemble approximately the
electromagnetic radius of the corresponding hadron while $\Delta z_h =
w/(\sqrt{2}\,m_h)$~\cite{Wirbel:1985ji} is fixed by the hadron mass
$m_h$ and the value $w = 0.35 - 0.5\,\GeV$ extracted from experimental
data. We adopt the values of $\Delta z_h$ and $S_h$ from our previous
work~\cite{Shoshi:2002in} as these values allow a good description of
many experimental data: $\Delta z_p = 0.3$ and $S_p = 0.86\,\fm$ for
protons, $\Delta z_{\pi} = 2$ and $S_{\pi} = 0.607\,\fm$ for pions,
and $\Delta z_{K} = 0.57$ and $S_{K} = 0.55\,\fm$ for kaons.

\subsection*{The Photon Wave Function}   

The photon wave function $\psi_{\gamma}(z_i,\vec{r}_i,Q^2)$ describes
the fluctuation of a photon with virtuality $Q^2$ into a
quark-antiquark pair with longitudinal quark momentum fraction $z_i$
and spatial transverse size and orientation $\vec{r}_i$. The
computation of the corresponding transition amplitude $\langle
q\qbar(z_i,\vec{r}_i)|\gamma(Q^2)\rangle$ can be performed
conveniently in light-cone perturbation theory~\cite{Bjorken:1971ah+X}
and leads to the following squared wave functions for transverse $(T)$
and longitudinally $(L)$ polarized photons~\cite{Nikolaev:1991ja}
\bea
\!\!\!\!\!\!\!\!\!\!\!\!\!|\psi_{\gamma_T^*}(z_i,\vec{r}_i,Q^2)|^2\! 
        &\!\!=\!\!&\!\frac{3\,\alphaEM}{2\,\pi^2} \sum_f e_f^2
                \left\{ 
                  \left[ z_i^2 \!+\! (1\!-\!z_i)^2\right]\epsilon_f^2\,K_1^2(\epsilon_f|\vec{r}_i|) 
                  + m_f^2\,K_0^2(\epsilon_f|\vec{r}_i|) 
                \right\}
        \label{Eq_photon_wave_function_T_squared} \\
\!\!\!\!\!\!\!\!\!\!\!\!\!|\psi_{\gamma_L^*}(z_i,\vec{r}_i,Q^2)|^2\! 
        &\!\!=\!\!&\!\frac{3\,\alphaEM}{2\,\pi^2} \sum_f e_f^2
                \left\{ 4\,Q^2\,z_i^2(1\!-\!z_i)^2\,K_0^2(\epsilon_f|\vec{r}_i|) \right\},
        \label{Eq_photon_wave_function_L_squared}
\eea
where $\alphaEM$ is the fine-structure constant, $e_f$ is the electric
charge of the quark with flavor $f$, and $K_0$ and $K_1$ are the modified
Bessel functions (McDonald functions). In the above expressions,
\be
        \epsilon_f^2 = z_i(1-z_i)\,Q^2 + m_f^2
\label{Eq_photon_extension_parameter}
\ee
controlls the transverse size(-distribution) of the emerging dipole,
$|\vec{r}_i| \propto 1/ \epsilon_f$, that depends on the quark flavor
through the current quark mass $m_f$. 

For small $Q^2$, the perturbatively derived wave functions,
(\ref{Eq_photon_wave_function_T_squared}) and
(\ref{Eq_photon_wave_function_L_squared}), are not appropriate since
the resulting large color-dipoles, i.e., $|\vec{r}_i| \propto 1/m_f
\gg 1\,\fm$, should encounter non-perturbative effects such as
confinement or chiral symmetry breaking. To take these effects into
account, we introduce $Q^2$-dependent quark masses, $m_f = m_f(Q^2)$,
that interpolate between the current quarks at large $Q^2$ and the
constituent quarks at small $Q^2$~\cite{Dosch:1998nw}. In our recent
work~\cite{Shoshi:2002in}, we adjusted the running quark masses as follows
\bea
        m_{u,d}(Q^2) 
        &=& 0.178\,\GeV\,(1-\frac{Q^2}{Q^2_{u,d}})\,\Theta(Q^2_{u,d}-Q^2) 
        \ , 
        \label{Eq_m_ud_(Q^2)}\\
        m_s(Q^2) 
        &=& 0.121\,\GeV + 0.129\,\GeV\,(1-\frac{Q^2}{Q^2_s})\,\Theta(Q^2_s-Q^2) 
        \label{Eq_m_s_(Q^2)}
        \ ,
\eea
with the parameters $Q^2_{u,d} = 1.05\,\GeV^2$ and $Q^2_s =
1.6\,\GeV^2$. For the charm quark mass, we used a fixed value of
\be
        m_c = 1.25\,\GeV
        \ .
\ee
%

\section{The Non-Forward \boldmath$T$-Matrix Element}
\label{App_T_tneq0}

In this appendix, we calculate the perturbative and non-confining
contribution to the non-forward ($t \neq 0$) $T$-matrix element. We
show explicitly that the non-forward $T$-matrix element depends on the
parameters which control the $z_i$\,-\,distribution of the wave
functions. These parameter are
essential for a good description of differential elastic cross
sections and the slope parameter as shown in our recent
work~\cite{Shoshi:2002in}.

The confining contribution to the non-forward ($t \neq 0$) $T$-matrix
element is not presented. It is much more complicated since some of
the integrations cannot be performed analytically. Nevertheless, we
find numerically that it shows the same features concerning the
$z_i$\,-\,distributions as the perturbative and non-confining
contributions.

The perturbative contribution to the non-forward $T$-matrix element in
the small-$\chi$
limit~(\ref{Eq_model_purely_imaginary_T_amplitude_small_chi_limit}),
\be
        T^{\pert}(s_0,t) 
        =\frac{2is_0}{9} \!\int \!\!d^2b_{\!\perp} 
        e^{i {\vec q}_{\!\perp} {\vec b}_{\!\perp}}
        \!\int \!\!dz_1 d^2r_1 \!\int \!\!dz_2 d^2r_2\,\,
        |\psi_1(z_1,\vec{r}_1)|^2 \,\,
        |\psi_2(z_2,\vec{r}_2)|^2\,\left(\chi^{\pert}\right)^2 \ ,
\ee
reduces upon integration over the impact parameter $|{\vec
  b}_{\!\perp}|$ to
\bea
        &&\!\!\!\!\!\!\!\!\!
        T^{\pert}(s_0,t) =
        \frac{32is_0}{9}
        \int d^2k_{\!\perp}\,\, 
        \alphaS(k^2_{\!\perp})\,\,
        i\tilde{D}_{\pert}^{\prime \,(2)}(k^2_{\!\perp})\,\,
        \alphaS\!\left(\!(\vec{k}_{\!\perp}+
        \vec{q}_{\!\perp})^2\!\right)\,
        i\tilde{D}_{\pert}^{\prime\,(2)}\!
        \left(\!(\vec{k}_{\!\perp}+\vec{q}_{\!\perp})^2\right)
        \nonumber \\
        &&\!\!\!\!\!\!\!\!\!\times
        \int_0^1\!\! dz_1 \int_0^1\!\! dz_2
        \left[H_1\left(z_1^2q^2_{\!\perp}\right)-
        H_1\left((z_1\vec{q}_{\!\perp}+
        \vec{k}_{\!\perp})^2\right)\right]\!\! 
        \left [H_2\left(z_2^2q^2_{\!\perp}\right)-
        H_2\left((z_2\vec{q}_{\!\perp}+
        \vec{k}_{\!\perp})^2\right)\right] 
        \nonumber \\
\label{tneq0_p}
\eea
with 
\be
H_i\left((z_i\vec{q}_{\perp}+\vec{k}_{\!\perp})\right) := 
     \int \!d^2r_i\,|\psi_i(z_i,\vec{r}_i)|^2\,
     e^{i\,\vec{r}_i\,(z_i\vec{q}_{\perp}+\vec{k}_{\!\perp})} \ . 
\label{H_def}
\ee
and $|\psi_i(z_i,\vec{r}_i)|^2$ denoting hadron or photon wave
functions.

The non-confining contribution to the non-forward $T$-matrix
element in the small-$\chi$ limit~(\ref{Eq_model_purely_imaginary_T_amplitude_small_chi_limit}),
\be
        T^{\nprt}_{nc}(s_0,t) 
        =\frac{2is_0}{9} \!\int \!\!d^2b_{\!\perp} 
        e^{i {\vec q}_{\!\perp} {\vec b}_{\!\perp}}
        \!\int \!\!dz_1 d^2r_1 \!\int \!\!dz_2 d^2r_2\,\,
        |\psi_1(z_1,\vec{r}_1)|^2 \,\,
        |\psi_2(z_2,\vec{r}_2)|^2\,\left(\chi^{\nprt}_{nc}\right)^2 \ ,
\ee
becomes analogously 
\bea
        &&\!\!\!\!\!\!\!\!\!
        T^{\nprt}_{nc}(s_0,t) =
        \frac{8is_0}{9}\left(\frac{\pi^2 G_2\,(1-\kappa)}{24}\right)^{\!\!2} 
        \!\!\int \!\frac{d^2k_{\!\perp}}{(2\pi)^2}\,\,
        i\tilde{D}_{1}^{\prime \,(2)}(k^2_{\!\perp})\,\,
        i\tilde{D}_{1}^{\prime\,(2)}
        \left((\vec{k}_{\!\perp}+\vec{q}_{\!\perp})^2\right)
\label{tneq0_nc}\\
        &&\!\!\!\!\!\!\!\!\!\times
        \int_0^1\!\! dz_1 \int_0^1\!\! dz_2
        \left[H_1\left(z_1^2q^2_{\!\perp}\right)-
        H_1\left((z_1\vec{q}_{\!\perp}+
        \vec{k}_{\!\perp})^2\right)\right]\!\! 
        \left [H_2\left(z_2^2q^2_{\!\perp}\right)-
        H_2\left((z_2\vec{q}_{\!\perp}+
        \vec{k}_{\!\perp})^2\right)\right]
        \nonumber 
\eea
with $H_{1,2}$ defined in~(\ref{H_def}). 

For $t = -q^2_{\perp} \neq 0$, both contributions~(\ref{tneq0_p})
and~(\ref{tneq0_nc}) depend on the shape of the $z_i$-distribution of
the wave function, i.e., for the Gaussian hadron wave
function~(\ref{Eq_hadron_wave_function}), the contributions depend on
the width $\Delta z_h$. This $\Delta z_h$-dependence is transferred to
the differential elastic cross section
\be
        \frac{d\sigma^{el}}{dt}(s,t) 
        = \inv{16 \pi s^2}|T(s,t)|^2
\label{Eq_dsigma_el_dt}
\ee
and its local slope
\be
        B(s,t) = 
        \frac{d}{dt} \left( \ln \left[ \frac{d\sigma^{el}}
        {dt}(s,t) \right] \right) \ .
\label{Eq_elastic_local_slope}
\ee
At $t=0$, the dependence on the shape of the $z_i$-distribution of the
wave functions disappears because of the normalization of the
$z_i$-distribution as can be seen immediately from~(\ref{H_def}).
Therefore, in our model the total cross sections -- related via the
optical theorem to the forward ($t = 0$) $T$-matrix element -- do not
depend on the parameter that characterize the $z_i$-distribution of
the wave function.

%
\end{appendix}

%
%
  
%
%
\end{document}